\documentstyle[12pt,psfig,subfigure,twoside]{article}

\begin{document}
\hbadness=10000
\pagenumbering {arabic}
\pagestyle{myheadings}
\markboth{J. Letessier, A. Tounsi, U. Heinz, J. Sollfrank and J.
Rafelski}          {Strangeness conservation in hot fireballs}

\title{\bf STRANGENESS CONSERVATION IN HOT FIREBALLS}
\author{$\ $\\ \bf Jean  Letessier \hspace{1.5cm}
\bf Ahmed Tounsi\\
Laboratoire de Physique Th\'eorique et Hautes Energies,
Paris\thanks{Unit\'e  associ\'ee au CNRS, Universit\'e PARIS 7, Tour 24,
5\`e \'et., 2 Place Jussieu, F-75251 CEDEX 05, France.}\\
$\ $\\ \bf Ulrich Heinz \hspace{1.8cm}
\bf Josef Sollfrank \\
Institut f\"ur Theoretische Physik, Universit\"at
Regensburg\thanks{Institut f\"ur Theoretische
Physik, Universit\"at Regensburg, Postfach 10 10 42,
W-8400 Regensburg, Germany} \\ $\ $\\
and\\$\ $\\
\bf Johann Rafelski \\
Department of Physics, University of Arizona, Tucson, AZ 85721}

\date{} 

\maketitle

\begin{abstract}

\noindent
A constraint between thermal fireball parameters arises from the
requirement that the balance of strangeness in a fireball is (nearly)
zero. We study the impact of this constraint on (multi-)strange
(anti-)baryon multiplicities and compare the hadron gas and
quark-gluon plasma predictions. We explore the relation between the
entropy content and particle multiplicities and show that the data are
compatible with the quark-gluon plasma hypothesis, but appear to be
inconsistent with the picture of an equilibrated hadron gas fireball.
We consider the implications of the results on the dynamics of evolution
and decay of the particle source.

\end{abstract}
\begin{center}
{\it Submitted to Physical Review D }
\end{center}
\vfill
{\bf PAR/LPTHE/92--27}\hfill\\
{\bf TPR-92-28}\hfill\\
{\bf AZPH-TH/92-23}\hfill{\bf November 1992}
\eject

\section{Introduction}
The reported enhancement of strange particle production in heavy ion
collisions relative to proton-proton and proton-nucleus collisions
\cite{strange} tells us that we are dealing with a state of matter
which is {\it i)} very dense or {\it ii)} relatively long-lived, or in
which {\it iii)} strangeness production cross sections are enhanced,
or which {\it iv)} possesses some combination of these three factors.
In order to be able to be more specific about the nature of the dense
matter, the relative production abundances of strange and
multi-strange baryons and anti-baryons were studied. They turned out
to be particularly sensitive probes of the thermal conditions of their
source \cite{MYSTR,Raf91}. Studies of the properties of a hadron
resonance gas fireball including strange particles have shown
\cite{RS88,ER91,LTR92,Cley92} that the constraint of vanishing net
strangeness of the emitting source leads to similar thermal and
chemical conditions for both a hadronic gas (HG) and a deconfined
quark-gluon plasma (QGP) fireball, if the temperature is taken to be
near $T=210\pm10$ MeV. This value was extracted \cite{Raf92} from the
$m_\bot$-spectra from 200 GeV\,A S--W and S--Pb collisions under the
assumption that their slope is dominated by the thermal motion of the
particles at freeze-out and does not contain a collective flow
component \cite{SH92}. We will discuss that the observed smallness of
the strange quark chemical potential appears to be inconsistent with
the presence of a strong flow component in the spectra.

It has already been pointed out \cite{LTR92} that near $T=210$ MeV
further information, such as the particle multiplicity per participant
nucleon, may help to distinguish QGP from the HG.  Following this
route, we will show here in detail \cite{Let92} that the data of
emulsion experiment EMU05 \cite{EMU05} indeed appear to be
incompatible on these grounds with the hadronic gas fireball model.
Further support for a new physical state such as the quark-gluon
plasma as the source of these anti-baryons may be obtained by studying
the response of the measured parameters to changes in beam energy or
size of the colliding nuclei. Our work has been in part motivated
by the imminent extension of the experimental program at CERN and BNL
to heavy nuclear projectiles (Pb and Au, respectively), and by the
prospect of higher (and lower) beam energies becoming available, in
particular towards the end of the decade at Brookhaven with RHIC and
possibly at CERN with LHC.  Therefore, in addition to $T\simeq 200$
MeV appropriate for the CERN--SPS energy range, we also include in our
discussion the parameter range ($T\simeq 150$ MeV) suitable for the
current BNL heavy-ion experiments at 10--15 GeV\,A and for potential
future low beam energy runs at CERN, as well as $T \simeq 300$ MeV,
which we judge appropriate for the future RHIC and LHC facilities
\cite{Aachen}.

Our expectation is that, as the conditions of the experiments are
changed, we will be able to study the response of the statistical
parameters such as the strange quark chemical potential $\mu_{\rm s}$
or the strangeness saturation factor $\gamma_{\rm s}$. These
parameters show characteristic differences in a hadron gas and a QGP:
$\gamma_{\rm s}$, for example, is expected to rapidly approach unity
for increasing four-volume of the QGP fireball, while remaining small
in HG fireballs \cite{KMR86}. The strange quark chemical potential
$\mu_{\rm s}$, on the other hand, has the characteristic property that
it is exactly zero in a strangeness-zero QGP fireball, independent of
its baryon density. In contrast to this, $\mu_{\rm s}$ is generally
different from zero in conventional hadronic gas reactions, except for
the special case of baryon-free matter with $\mu_{\rm B}=0$; however,
if $T$ is not too large ($T<220$ MeV), for each value of the
temperature a certain value of baryochemical potential $\mu_{\rm B}
\ne 0$ can be found for which also $\mu_{\rm s}^{\rm HG}=0$. This
value generally depends on the detailed form of the equation of state
(EoS), and for the case of a conventional (Hagedorn type) HG it has
been discussed in Refs. \cite{LTR92,Raf87}. While the large values of
$\gamma_{\rm s}$ which are characteristic of the fast chemical
kinetics in a QGP are largely preserved during hadronization, the
value of $\mu_{\rm s}$ is usually subject to modifications, in
particular if the phase transformation occurs slowly. Experimentally,
the values of $\mu_{\rm s}$ and $\gamma_{\rm s}$ are reflected in the
observed strange hadron ratios (where, as we will discuss, in the case
of QGP certain model assumptions for the hadronization process are
necessary in order to quantify this relationship), which allows for a
rather straightforward experimental test of the fireball picture, of
the concept of (absolute or relative) chemical equilibrium, and of the
nature (HG or QGP) of the collision fireball.

In order to distinguish between the two phases by studying particle
production as the conditions of the experiments are changed we need to
understand also in the HG phase how the strange chemical potential
$\mu_{\rm s}$ relates to the baryochemical potential $\mu_{\rm B}$ in
the collision fireball. Such a relationship is required by the fact
that the total strangeness in the initial state is zero, and that
strangeness is conserved by strong interactions. Thus, up to
(presumably small) pre-equilibrium emission effects which
can introduce a small strangeness asymmetry $\bar s-s\equiv\varepsilon
s$, the number of strange and anti-strange quarks in the fireball are
equal ($\varepsilon\simeq 0$). We work out in detail how the strange
chemical potential is determined at fixed temperature by the
baryochemical potential and a given (small) fraction $\varepsilon$ of
net strangeness in the fireball. We find that at a fireball
temperature around $T= 210$ MeV, there is a peculiar behavior of
$\mu_{\rm s}$ which causes a particular insensitivity of the strange
anti-baryon ratios alone to the structure of the source. In order to
be able to reach a conclusion about the internal structure of the
source, we then study the entropy content of the particle source. We
also show that the kaon to hyperon ratio is a poor signature for the
nature of the source, but could be a good measure of the baryochemical
potential \cite{KRG83}.

The HG part of our work is similar in spirit, but different in detail
from the parallel effort of Cleymans and Satz
\cite{Cley92}\footnote{When comparing the present work with
Ref.~\cite{Cley92}, it should be noted that these authors use as the
strangeness potential the chemical potential $\mu_{\rm S}$ of the
(anti-)hyperon number, while we prefer to work directly with the
strange quark chemical potential $\mu_{\rm s}$. The relationship
between their $\mu_{\rm S}$ and our $\mu_{\rm s}$ is given by
$\mu_{\rm s} = \mu_{\rm B}/3 - \mu_{\rm S}$.\label{foot1}}. We study
particle ratios at fixed transverse mass $m_\bot>1.7$ GeV. We allow
for the strangeness phase space to be only partially saturated and
show how the degree of saturation $\gamma_{\rm s}$ can be
experimentally measured in order to later serve as a test for the
kinetic theory of strangeness production. We allow the strangeness to
be slightly unbalanced (up to 10\%) and discuss the corresponding
variations in the observables. Furthermore we compare the entropy
content for the two different scenarios (HG and QGP) and confront the
findings with the observed particle multiplicity. In particular we
consider the ratio of net charge to total charged multiplicity,
$D_{\rm Q} \equiv (N^+ - N^-)/(N^+ + N^-)$, which can be measured
(without identifying particles) with any tracking device within a
magnetic field; we show that this ratio is directly related to the
specific entropy $\cal S/B$ of the fireball, which should be
considerably larger for a QGP than for a HG \cite{LTR92}. Since, due
to the different stopping, the experimental results from S--S and S--W
collisions at 200 GeV\,A are not in the same class, we avoid combining
these results because this would weaken the conclusions from our
analysis. We do not combine in our analysis the global kaon data with
anti-hyperon data, since low $p_\bot$ kaons, unlike strange
anti-baryons, can also arise from peripheral, spectator related
processes and other rescattering processes and are therefore not
necessarily messengers from the same stages of the collision. Instead
we consider the $\Lambda/p$ and $\overline\Lambda/\overline p$ as well
as $\Omega/\Xi^- $ and $\overline{\Omega}/\overline{\Xi^- }$ ratios
which, as we will explain, provide independent tests of the degree of
strangeness saturation~$\gamma_{\rm s}$.

At this point it is worth recording that if a QGP state was formed (as
may be indicated by the relatively large value of $\gamma_{\rm s}$, see
below), the strangeness pair abundance per participating baryon in this
state would be about a factor 2.5 larger compared to hadronic gas
interactions. On the other hand, there would be a similar enhancement in
the produced entropy (which will manifest itself in the total
multiplicity), resulting in a difficulty to identify the absolute
enhancement of both quantities: the simplest observable, the $K/\pi$
ratio, is therefore not a very good discriminator in this context
\cite{GR85}. As we will show, measurement of the specific entropy $\cal
S/B$, either globally via the total particle multiplicity per
participating nucleon, or (better, see below) in the same kinematic range
where the produced strange particles are observed via the ratio $D_{\rm
Q}=(N^+ - N^-)/(N^+ + N^-)$, can remove this ambiguity.

One of the important objectives we pursue is to present a large number
of predictions for (strange) particle production which are capable to
test the internal consistency of the model developed here and can shed
new light on the problem of distinguishing the HG and QGP fireballs.
These predictions are quantitative for the S--W collision system at
200~GeV\,A, but are of a more qualitative nature for the yet
unexplored Pb--Pb collisions where we cannot predict precisely the
governing statistical parameters (temperature, chemical potentials).
We also discuss the systematic behavior of the different statistical
model observables in changing environments.

Perhaps the most notable conclusion we reach concerns the model of
hadronization of the fireball: we conclude that the only consistent
model of the reaction involves explosive disintegration of a QGP as
the source of the observed strange anti-baryons. This allows the
chemical potentials of the quarks in the QGP phase to get transferred
without modification to the observed high-$m_\perp$ strange baryons
and anti-baryons, via quark coalescense, and explains naturally the
observed vanishing of the strange chemical potential.
Since only high $m_\perp$ domain of strange anti-baryon spectra has been
experimentally explored, the simple picture of a complete explosion is not
absolutely necessary.
It is conceivable that at low $m_\perp$, in particular for low-$p_\perp$
pions, another slower hadronization mechanism also plays a
role. We do see the need for some mechanism that produces additional
pions, in excess over those obtained from the explosive quark
coalescence picture, in order to save the entropy balance and to account
for the relatively large observed multiplicity densities. These
excess pions (plus perhaps some other light mesons) could easily arise
from the gluons in the QGP which carry a large fraction of its
entropy.
A quantitative formulation of such a hadronization model,
however, is not yet available and would require additional data on the
detailed origin of the excess multiplicity and its distribution in
momentum space.

Naturally, these conclusions rest on the assumed validity of the
thermal fireball model. While at present the number of parameters just
equals the number of independent observables, the model has
considerable predictive power and, with the help of the predictions
presented in this paper, its validity will be (hopefully soon) tested.
Other pictures for the source of the observed hadrons, such as an
equilibrated HG or a QGP fireball which evolves slowly and in full
equilibrium into a HG fireball, can not account consistently for
certain important experimental features of strange anti-baryon
production.

In section \ref{sect2} we present our thermal model with
special attention given to the parameters describing the chemical
equilibrium. We take into account the $u$-$d$ asymmetry and study the
condition of stangeness balance. In section \ref{sect3} we determine
the thermodynamical parameters using the WA85 data on strange baryon
and anti-baryon production (with and without taking into account
resonance decays). In section \ref{sect4} the implication of these
parameters on the entropy content of the QGP and HG fireballs and on
the multiplicity are discussed. We then confront our model with EMU05
and NA35 data. In section \ref{sect5} we discuss the
implications of our results regarding the nature of the initial
fireball and the dynamic of its evolution. We give some concluding
remarks concerning the distinction we draw between QGP and HG.

\section{Thermal models}\label{sect2}

The use of thermal models to interpret data on particle abundances and
spectra from nuclear collisions is motivated by the hope and expectation
that in collisions between sufficiently large nuclei at sufficiently
high energies a state of excited nuclear matter close to local
thermodynamic equilibrium can be formed, allowing us to study the
thermodynamics of QCD and the possible phase transition from a HG to a
QGP at a critical energy density. Since we do not yet reliably know
how big the collision system and how large the beam energy has to be
for this to occur (if at all), thermal models should be considered as
a phenomenological tool to test for such a behavior. Even if
successful, the validity of such an approach must be checked later by
a more detailed theoretical analysis of the kinetic evolution of the
collision fireball.

It has been noted repeatedly that, especially in collisions involving
heavy nuclear targets, the observed particle spectra, after correcting
for resonance decay effects \cite{sollfrank}, resemble thermal
distributions with a common temperature $T$ (although a flow component
cannot presently be excluded and is actually quite likely
\cite{SH92}), and that so far the interpretation in terms of a
generalized thermal model, which assumes equilibration only of the
momentum distributions and the light quark abundances, but allows for
deviations in particle abundances from {\it absolute chemical}
equilibrium, in particular in the strange sector, appears to  be
consistent with the observed particle ratios \cite{MYSTR}. In this
section we will discuss this model as far as it is relevant for our
analysis.

\subsection{Fireball parameters}
\label{2.0N}

Let us assume that in collisions between two nuclei a region with
nearly thermal equilibrium conditions is formed near ``central''
rapidity, {\it i.e.} at rest in the center of momentum frame of the
projectile and a target tube with the diameter of the projectile.

This central fireball from which the observed particles emerge is
described by its temperature $T$ and by the chemical potentials
$\mu_i$ of the different conserved quark flavors $u,d,s$. Since the
conserved quantum numbers of hadrons are simply the sum of the
corresponding quantum numbers of their quark constituents, the
hadronic chemical potentials are given as the sum of constituent quark
chemical potentials. Since the strong and electromagnetic interactions
do not mix quark flavors, $u,d,$ and $s$ quarks are separately
conserved on the time scale of hadronic collisions, and they can only
be produced (or annihilated) in pairs: $X\leftrightarrow q_i\bar q_i$.
This implies that in {\it absolute chemical equilibrium} the chemical
potentials for particle and anti-particle flavors are opposite to each
other. This observation can be restated in terms of the particle
fugacities which are convenient for counting particles and are related
to the chemical potentials by:
 \begin{equation}
   \lambda_i = e^{\mu_i/T}\ .
 \end{equation}
Since the quantum numbers of a hadron are simply the sum of the
corresponding quantum numbers of their constituent quarks, the fugacity of
each HG species is simply the product of the quark fugacities, {\it
viz.}\/ $\lambda_{\rm p} = \lambda_{\rm u}^2\lambda_{\rm d},\
\lambda_{{\rm K}^+}=\lambda_{\rm u} \lambda_{\bar {\rm s}}$, {etc}.
Consequently the relation between particle and anti-particle fugacity
is in general:
 \begin{equation}
  \lambda_{\bar i}=\lambda_i^{-1}\, .
  \label{lam}
 \end{equation}

If absolute equilibrium is not reached, this relationship can still be
maintained by introducing an additional parameter characterizing the
approach to equilibrium. For strange particles, specifically, absolute
chemical equilibrium is not easy to obtain: due to the higher mass
threshold, the production of $\bar ss$ pairs usually proceeds much
more slowly than that of $\bar uu$ and $\bar dd$ pairs. This is true
both in hadronic interactions and in the quark-gluon plasma
\cite{KMR86}. In fact, the $s\bar s$ equilibration time scale is
estimated to be of the same order of magnitude as the total lifetime
of the collision fireball in the QGP phase \cite{RM82}, and much
longer than that in the hadronic phase \cite{KR85}. On the other hand,
once a certain number of strange-anti-strange pairs has been created
in such a hadron gas, their redistribution among the various strange
hadronic species is no longer hindered by any thresholds and occurs
relatively fast. It thus makes sense to introduce the concept of {\it
relative chemical equilibrium}, in which the strange phase space is
not fully saturated, but whatever strangeness has already been created
is distributed among the available strange hadron channels according
to the law of maximum entropy. Parametrizing the approach to strange
phase space saturation by a factor $0 < \gamma_{\rm s} \leq 1$, the
individual hadron fugacities are then given by
 \begin{equation}
   \gamma_i \lambda_i \equiv \gamma_{\rm s}^{n(i,s)}\, \lambda_i \, ,
 \label{gammai}
 \end{equation}
where the $\lambda_i$ still satisfy the chemical equilibrium condition
(\ref{lam}), and where the power $n(i,s)$ counts the total number of
strange valence quarks {\it plus} anti-quarks in hadron species $i$.

Altogether, within the thermal model the fireball is thus described by  5
parameters: its temperature $T$, its baryon chemical potential
 \begin{equation}
  \mu_{\rm B}=3\mu_{\rm q}\, ,
 \end{equation}
the strangeness chemical potential $\mu_{\rm s}$, the light quark
asymmetry potential $\delta\mu$ (see subsection~\ref{2.1}) and the
strangeness saturation factor $\gamma_{\rm s}$. The value of the
temperature can be  extracted from the observed transverse momentum
spectra whose exponential slope  is governed by a combination of
thermal motion and collective expansion of the fireball \cite{SH92}.
The separation of these two  contributions, {\it i.e.} the isolation of
$T$, is not straightforward, but has been studied extensively in
\cite{SH92}. Given $T$, the number of independent variables is
further reduced by studying the $u$--$d$ asymmetry in the fireball  and
relating it to the asymmetry potential $\delta\mu$, and by the
condition of strangeness conservation. This leaves us with two
independent variables, say $\mu_{\rm B}$ and $\gamma_{\rm s}$, which
have to be determined from measured particle ratios.

The various parameters described above enter the thermodynamic
description through the partition function from which all observables
are derived. For the hadronic gas state we follow here the conventional
hadronic gas model first developed by Hagedorn \cite{HAG78} in which the
equation of state of any hadronic system is obtained from a partition
function which sums over all hadronic resonances. It is believed that by
including a complete set of intermediate scattering states in the form of
(usually zero-width) hadronic resonances the dominant interactions within
the hadronic system are automatically taken into account. However, an
important but tacit approximation is made in this approach, concerning
the parameters such as mass (and sometimes widths) of the resonances
which are inserted with their free space values. This neglects the
possibility that in particular at high temperature and particle or
baryon density, there may occur significant shifts from the free space
values--one often speaks in this context of the melting of hadronic
masses as one approaches extreme conditions. While we do our
computations entirely within the conservative framework of the
conventional model, our general physics discussion will be shaped in
such a way that our conclusions should not critically depend on the
exact validity of the model used.

The standard hadronic gas model is certainly not sufficiently precise
for our purpose if we were to ask questions concerning the absolute
values of energy, entropy, baryon density, etc. which depend either on
the ever increasing mass spectrum of particles or on the proper
volume occupied by the particles \cite{TOUNSI91}. However, the
condition of zero strangeness and quantities such as the entropy per
baryon, which involve ratios of extensive variables, are independent  of
the absolute normalization of the volume and of the renormalization
introduced by the diverging particle spectrum and hence can safely be
considered in our approach.

We now turn to the discussion of the constraints which limit the
number of the free thermal parameters in the fireball.

\subsection{{\it u--d} asymmetry}
\label{2.1}

In our calculations we will distinguish between the $u$ and $d$ quarks
by introducing separate fugacities, $\lambda_{\rm u}$ and $\lambda_{\rm
d}$, instead of $\lambda_{\rm q}$ in the partition function. Since  even
in the heaviest nuclei there is only a small $u$--$d$ asymmetry due  to
the neutron excess, we find it convenient to work with the
variables:
 \begin{eqnarray}
  \mu_{\rm q}&&\hspace{-0.6cm}=(\mu_{\rm d}+\mu_{\rm u})/2 \ ,\\
  \delta\mu&&\hspace{-0.6cm} =\mu_{\rm d}-\mu_{\rm u} \ .
 \label{eq3}
 \end{eqnarray}
Here, $\mu_{\rm q}$ is the ``light quark'' chemical potential ($\mu_{\rm
B}/3$ in terms of the baryochemical potential) and $\delta\mu$ describes
the asymmetry in the number of up and down quarks due to the neutron
excess in the heavy ion collision. The ratio of the net number of down
and up quarks in the fireball is
 \begin{eqnarray}
   R_{\rm f}={\langle d-\bar d\rangle
        \over \langle u-\bar u\rangle}\, .
 \label{eq4}
 \end{eqnarray}
In a central S--W collisions, where a tube with the transverse area of
the S projectile is swept out from the W target and participates in the
fireball $R_{\rm f}^{\rm S-W}\simeq1.08$; in Pb--Pb collisions
$R_{\rm f}^{\rm Pb-Pb}=1.15$. The value of $\delta\mu$ is at each fixed
$T$ determined by the value of $R_{\rm f}$, but depends on the assumed
structure of the source, {\it i.e.} the EoS. We have investigated this
relation for the two cases considered here: the conventional HG and the
QGP.

\pagebreak[5]

$\ $

\vspace{-1cm}

\begin{figure}[t]
\begin{minipage}[t]{0.475\textwidth}
\hspace{0.5cm}
In the HG phase we include in the partition function all non-strange
mesons up to a mass of 1690 MeV, all nucleons  up to 1675 MeV and all
$\Delta$'s up to 1900 MeV. [We note that higher resonances would matter
only if their number were divergent (as is the case in the Bootstrap
approach of Hagedorn \cite{HAG78}) and if the HG  was sufficiently long
lived to populate all high mass resonances]. In addition strange hadrons
are included as described in Eq.\,(\ref{4a}) below, which we expand in
calculations to allow to distinguish between $u$ and $d$ quarks within
the various strange hadron multiplets.

\hspace{0.5cm}
In Fig.\,\ref{F1} we show how in the conventional hadron gas the
difference $\delta\mu / \mu_{\rm q}$ induced by the isospin asymmetry of
the fireball depends on the temperature, for two selected values $R_{\rm
f}=1.08$ and $1.15$ (appropriate for S--W and Pb--Pb collisions,
respectively), and for three fixed values of $\lambda_{\rm s} = 0.95,
1.00,$ and 1.05.
\end{minipage}\hfill
\begin{minipage}[t]{0.475\textwidth}\vspace{-2cm}
\centerline{\hspace{0.2cm}\psfig{figure=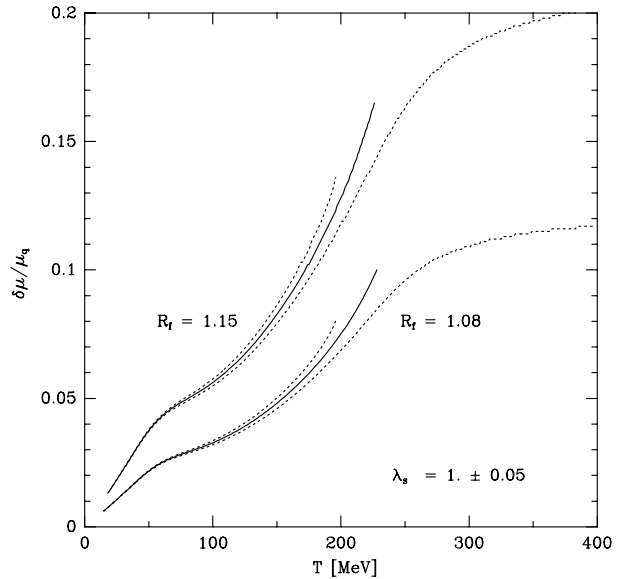,height=11.5cm}}
\vspace{ -2.3cm}
\caption{\tenrm
$\delta\mu/\mu_{\rm q}$ asymmetry for $R_{\rm f} = 1.08$ (S--W
collisions) and $R_{\rm f} = 1.15$  (Pb--Pb collisions).  Solid lines:
$\lambda_{\rm s} = 1.00$; upper dotted lines: $\lambda_{\rm s} = 0.95$;
lower  dotted lines: $\lambda_{\rm s} =1.05$. \protect\label{F1}}
\end{minipage}
\end{figure}
$\ $

\vspace{-1.1cm}
 $\ $\hspace{3.8cm} The curves shown correspond to non-trivial solutions
({\it i.e.} solutions with $\mu_{\rm q}\ne 0$) of the strangeness
balance equation with the given values for $\lambda_{\rm s}$; as we
will explain below, these solutions cease to exist (since we require
$\langle s-\bar s \rangle \simeq 0$) above a certain critical
temperature, which is given by $T\simeq 230$ MeV for $\lambda_{\rm
s}=1$ and smaller (larger) for smaller (larger) values of
$\lambda_{\rm s}$. We see that, while for $T\to 0$ the ratio
$\delta\mu / \mu_{\rm q}$ vanishes in the hadron gas, it is of the
same order as the deviation from unity of $R_{\rm f}$ once $T$ reaches
values around 200 MeV which are of interest here.

For the QGP the ratio $\delta\mu / \mu_{\rm q}$ is independent of
$\lambda_{\rm s}$, due to the decoupling of the strange and non-strange
chemical potentials in the partition function. For $\mu_{\rm q}<\pi T$
one finds \cite{Raf91}
 \begin{eqnarray}
   R_{\rm f}^{\rm QGP} \simeq {\mu_{\rm d}\over\mu_{\rm u}}
                       \simeq 1+{\delta\mu\over \mu_{\rm q}}\, ,
 \end{eqnarray}
such that $\delta\mu / \mu_{\rm q}$ is exactly equal to $R_{\rm f}-1$.
It is remarkable that towards the maximum temperature consistent with
strangeness conservation the HG and QGP based results for $\delta\mu$ at
a given $R_{\rm f}$ nearly agree. We thus find that, although small,  the
difference between the chemical potentials of $u$ and $d$ quarks  is not
always negligible. We note for later applications that in the region of
interest to us here ($T\sim 210$ MeV) we have irrespective of the nature
of the fireball:
 \begin{equation}
   {\delta\mu\over T} \approx {\mu_{\rm q}\over T}(R_{\rm f}-1)\ .
\label{deltamu}
 \end{equation}

\subsection{The strangeness partition function}
\label{2.2}

In the Boltzmann approximation, it is easy to write down  the partition
function for the strange particle fraction of the hadronic gas, ${\cal
Z}_{\rm s}$, and we follow the notation of Ref.\,\cite{Raf87}. However,
given the recently measured values of  the thermal parameters (which
allow in particular for a rather high temperature $T\sim 210$ MeV) it is
necessary to sum over some more strange  hadronic particles than was done
in Eq.\,(\ref{eq4}) of Ref.~\cite{Raf87}. Including the possibility of
only partially saturated strange phase space through the factor
$\gamma_{\rm s}$  above, but neglecting for the moment the isospin
asymmetry $\delta \mu$, we have:
 \begin{eqnarray}
   \ln{\cal Z}_{\rm s}^{\rm HG} = { {V_{\rm h}T^3}
     \over {2\pi^2}} \!\!&\Bigl[&\!\!\!
     (\lambda_{\rm s} \lambda_{\rm q}^{-1} +
     \lambda_{\rm s}^{-1} \lambda_{\rm q}) \gamma_{\rm s} F_K +
     (\lambda_{\rm s} \lambda_{\rm q}^{2} +
     \lambda_{\rm s}^{-1} \lambda_{\rm q}^{-2}) \gamma_{\rm s} F_Y
  \nonumber \\&&\!\! +
   (\lambda_{\rm s}^2 \lambda_{\rm q} + \lambda_{\rm s}^{-2}
     \lambda_{\rm q}^{-1}) \gamma_{\rm s}^2 F_\Xi +
    (\lambda_{\rm s}^{3} + \lambda_{\rm s}^{-3})
     \gamma_{\rm s}^3 F_\Omega \,     \Bigr]
  \label{4a}
 \end{eqnarray}
where the kaon ($K$), hyperon ($Y$), cascade ($\Xi$) and omega ($\Omega$)
degrees of freedom in the hadronic gas are included successively. The
phase space factors $F_i$ of the various strange particles are:
\begin{eqnarray}
   F_K &&\hspace{-0.6cm}=  \sum_j g_{K_j} W(m_{K_j}/T);\
           K_j=K,K^\ast,K_2^\ast,\ldots ,\ \
           m\le 1650 \ {\rm MeV}\, ,
   \nonumber\\
   F_Y &&\hspace{-0.6cm}=  \sum_j g_{Y_j} W(m_{Y_j}/T);\
           Y_j=\Lambda, \Sigma, \Sigma(1385),\ldots ,\ \
           m\le 1750\ {\rm MeV}\, ,
   \nonumber\\
   F_\Xi &&\hspace{-0.6cm}=  \sum_j g_{\Xi_j} W(m_{\Xi_j}/T);\
           \Xi_j=\Xi,\Xi(1530),\ldots ,\ \
           m\le 1820\ {\rm MeV}\, ,
   \nonumber\\
   F_\Omega &&\hspace{-0.6cm}=  \sum_jg_{\Omega_j} W(m_{\Omega_j}/T);\
       \Omega_j=\Omega .
 \end{eqnarray}
where the $g_i$ are the spin-isospin degeneracy factors,
$W(x)=x^2K_2(x)$, and $K_2$ is the modified Bessel function.

For the quark-gluon plasma, on the other hand, the strange
contribution to the partition function is much simpler, because the
strange quarks and anti-quarks are isolated and do not occur in
clusters with non-strange quarks as in the hadron gas:
 \begin{equation}
   \ln{\cal Z}_{\rm s}^{\rm QGP} =
    {{g_{\rm s}\,V\,T^3} \over {2\pi^2}}
    \int p^2\, dp\, \left[ \ln
    \left( 1 + \gamma_{\rm s}\lambda_{\rm s}
               e^{-\sqrt{m_{\rm s}^2+p^2}/T} \right) +
  \ln   \left( 1 + \gamma_{\rm s}\lambda_{\rm s}^{-1}
               e^{-\sqrt{m_{\rm s}^2+p^2}/T} \right) \right]\, .
 \label{lnZs}
 \end{equation}
with the strange quark spin-color degeneracy factor
 \begin{equation}
   g_{\rm s} = 2 \times 3 = 6\, .
 \label{gs}
 \end{equation}
For the value of the strange quark mass we take $m_{\rm s} \sim$ 150--180
MeV; in Eq.\,(\ref{lnZs}) $\gamma_{\rm s}$ and $\lambda_{\rm s}$ do not
factorize as has been the case for the hadron gas, because $T$ is of
order $m_{\rm s}$ and the Boltzmann approximation is not always
satisfactory and can  produce errors of the order of 20\%. Recall that in
general the strange particle density in a QGP is larger by a factor  2--5
than in  a HG, due to the lower threshold ($m_{\rm s}<m_{\rm K}$) and the
presence of the color degeneracy factor in Eq.\,(\ref{gs}).

\subsection{Strangeness balance}
\label{2.3}

We will now discuss the reduction of the number of free parameters
caused by the condition of strangeness balance. Since the net
strangeness of the colliding nuclei in the initial state is zero, and
strangeness is conserved by strong interactions, the net strangeness of
the collision fireball will stay close to zero throughout the
collision, except for corrections due to possible strangeness
asymmetry in  pre-equilibrium and asymmetric surface radiation of
strange hadrons \cite{CGrei}. If these processes can be neglected, the
resulting condition of vanishing total strangeness takes the form
\begin{equation}
    0 = \rho_{\rm s} = \langle s \rangle - \langle \bar s \rangle
     = \lambda_{\rm s} { \partial \over {\partial \lambda_{\rm s}}}
                   \ln {\cal Z}_{\rm s} \, ,
 \label{Eq6}
 \end{equation}
where $\rho_{\rm s}$ is the net strangeness density,
 \begin{equation}
   \rho_{\rm s} = \sum_i s_i \rho_i\, ,
 \label{rhos}
 \end{equation}
with $s_i$ being the strangeness of particle species $i$ having
density $\rho_i$, and the sum going over all particle species $i$
existing in the respective phase (QGP or HG). While for the QGP this
relation leads always to the result $\lambda_{\rm s}=1$ or $\mu_{\rm
s}=0$ (as one easily verifies by inserting Eq.\,(\ref{lnZs})), for the
hadron gas it is an implicit equation relating $\lambda_{\rm s}$ to
$\lambda_{\rm q}$ in a way which depends on the temperature $T$
\cite{Raf87,LRH88}.

In order to also be able to account for  pre-equilibrium and surface
emission fluctuations, we introduce the net strangeness fraction
 \begin{equation}
   \varepsilon \equiv {\langle \bar s \rangle - \langle s \rangle
   \over \langle s \rangle}\ ,
 \end{equation}
and will also consider solutions for the analogue of Eq.\,(\ref{Eq6}),
where the left hand side is given by $-\varepsilon \langle s\rangle$. In
this work we restrict ourselves to asymmetries $\vert \varepsilon  \vert
\leq 0.1$ which might be accessible by purely statistical
fluctuations caused by surface emission, without ``strangeness
distillation" during QGP hadronization \cite{LRH88,CGrei} which arises
when chemical equilibrium is enforced during a slow QGP hadronization
process. An analysis of the phase diagram and the strangeness balance
conditions for systems with very large net strangeness is presented
elsewhere \cite{LH92}. We note briefly the practical consequences of
$\varepsilon = \pm 0.1$: for the fireball created in central S--W
collisions one has about 108 baryons in the interaction tube. If we
consider a hadron gas with the thermal parameters (without flow
component) as determined for these S--W collision data \cite{LTR92},  we
find a strange pair abundance of about 0.4 per baryon or about 40
strange pairs in total. The maximum asymmetry considered here in the  HG
phase at $T\simeq 200$ MeV is thus about $\pm 4$ strange quarks.
Fluctuations of this magnitude are expected to occur in about half of
the collision systems.

\subsubsection{The case $\mu_{\rm s}=0$ ($\lambda_{\rm s}=1$)}
\label{2.3.1}

Since the vanishing of $\mu_{\rm s}$ for all values of $T$ and
$\mu_{\rm B}$ in a strangeness neutral QGP appears to be such a unique
feature of this state, it is interesting to ask to what extent this
condition could arise in a HG. We will therefore first study the
mechanisms which lead in a HG model to a small strange quark chemical
potential in a restricted region of temperatures. Of course, for a
baryon-free system ($\mu_{\rm B}=0$) even in the hadron gas
$\mu_{\rm s}=0$ is a natural consequence of strangeness neutrality. For
systems with non-vanishing baryon number (like the central
rapidity fireballs created at AGS and SPS energies) this is no
longer true. We therefore ask ourselves under which conditions
$\mu_{\rm s}=0$, {\it viz.}\/ $\lambda_{\rm s}=1$, allows for non-zero
values of $\mu_{\rm B}^0$ to be consistent with strangeness
neutrality. This question can be answered analytically: inserting the
partition function (\ref{4a}) into the strangeness neutrality
condition (\ref{Eq6}), we see that at $\lambda_{\rm s}=1$ the phase
space for $\Omega$ and $\overline{\Omega}$ cancels out, giving the  exact
result
 \begin{eqnarray}
  \mu_{\rm B}^0=3\,{\rm cosh}^{-1}\left({F_{\rm K}\over 2F_{\rm Y}}
-\gamma_{\rm s} {F_{\Xi}\over F_{\rm Y}}\right)\, .
 \label{muB}
 \end{eqnarray}
There is a real solution only when the argument on the right hand side
is larger than unity. It turns out that this condition is a sensitive
function of the temperature $T$ and of the hadronic resonances
included (and therefore of the masses of the resonances). In
particular, for any given resonance mass spectrum used to compute the
phase space factors $F_i$, there is a temperature $T_0$ beyond which  no
such solution is possible --- this occurs since $F_{\rm K}/F_{\rm  Y}$ is
a  monotonically decreasing function of $T$ (see Fig.\,1 in
Ref.\,\cite{Raf87}).

The case of exact strangeness neutrality ($\varepsilon=0$) in a HG in
absolute chemical equilibrium ($\gamma_{\rm s}=1$) was considered in
\cite{Cley92}, and for this choice of parameters our following
theoretical analysis essentially agrees with that work.

\subsubsection{HG behavior of $\mu_{\rm s}$ and $\mu_{\rm B}$ for nearly
conserved strangeness}
\label{2.5}
An important aspect to consider is the value of $\mu_{\rm s}$ observed
and how it is compatible with the picture of the reaction. The reason
that $\mu_{\rm s}$ is of such importance is the expected value  $\mu_{\rm
s}=0$ associated with QGP. We first will address the question under what
circumstances this value is accessible to a HG fireball, and then turn to
discuss more generally how different dynamical scenarios can be
distinguished. We expect that there is a domain of  temperatures for
which even a considerable change of $\mu_{\rm B}$  between $\mu_{\rm
B}=0$ and $\mu_{\rm B}^0$ does not induce a  significant change of
$\mu_{\rm s}$.

\begin{figure}[ht]\vspace{-4cm}
\begin{minipage}[t]{0.475\textwidth}
\centerline{\hspace{0.2cm}\psfig{figure=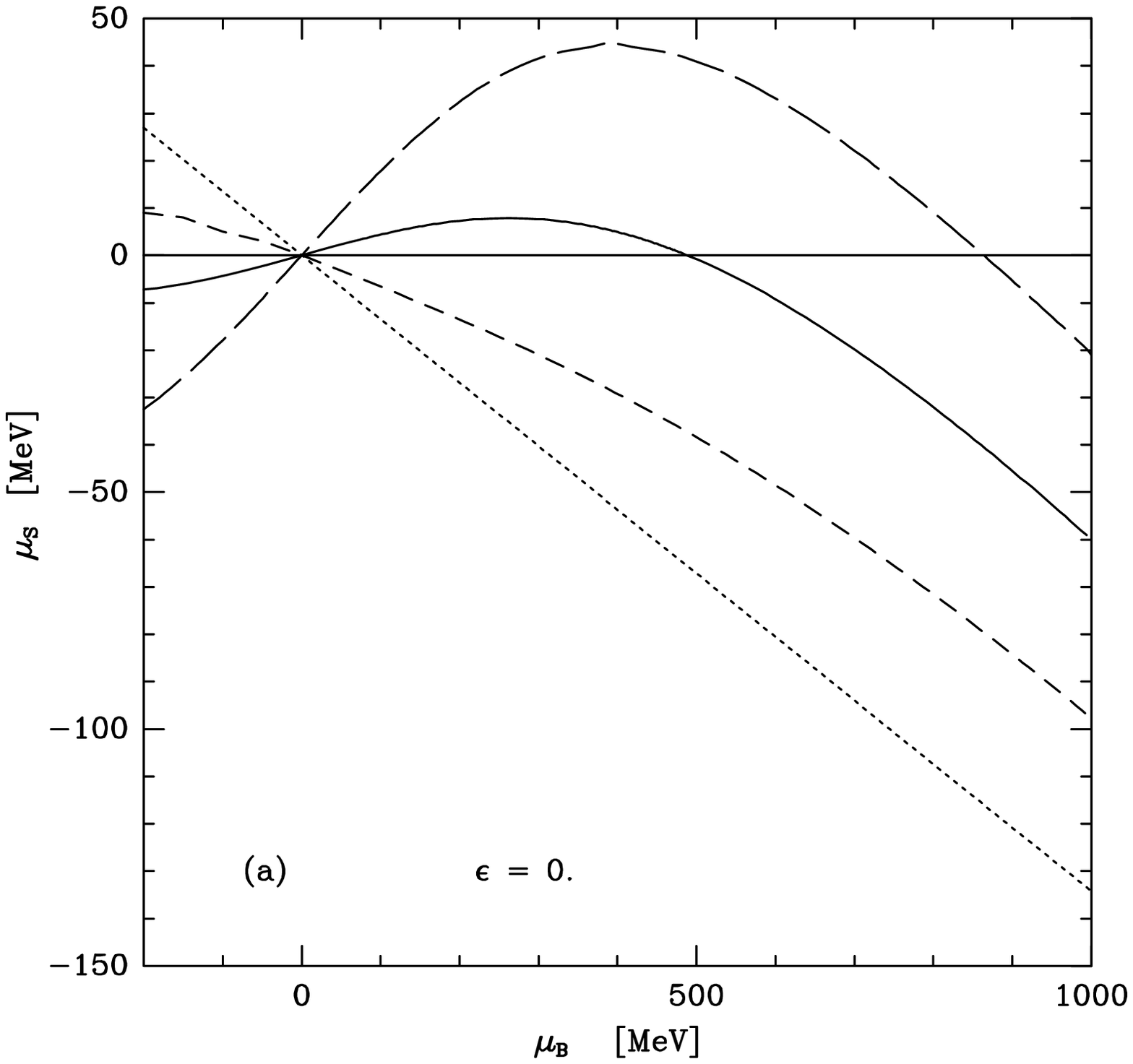,height=11.5cm}}\vspace{-2cm}
\end{minipage}\hfill
\begin{minipage}[t]{0.475\textwidth}
\centerline{\hspace{0.2cm}\psfig{figure=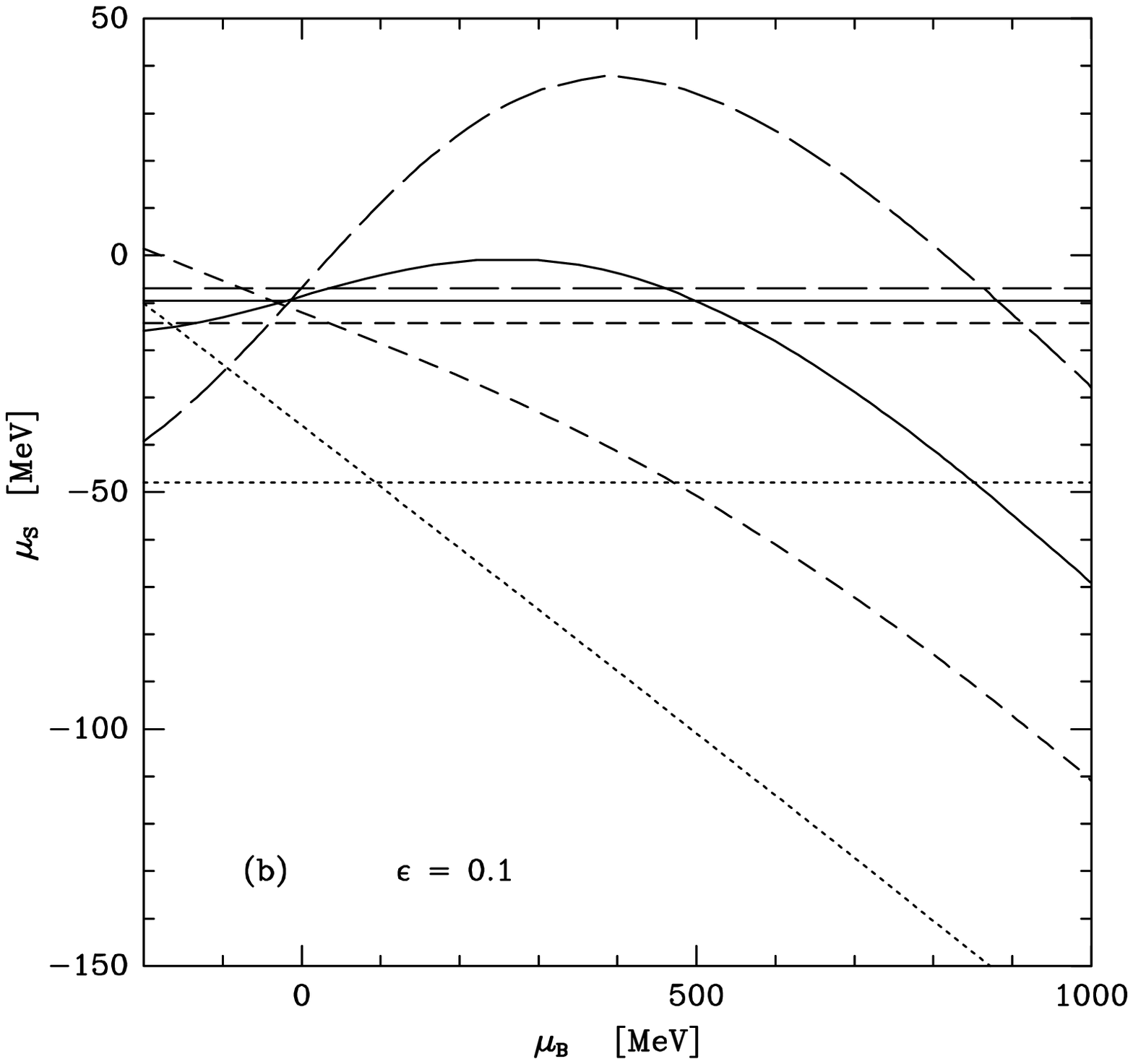,height=11.5cm}}\vspace{-2cm}
\end{minipage}\\
\begin{minipage}[t]{0.475\textwidth}
\vspace{-1.cm}
\caption[s-chem.pot vs. baryo. chem. pot]
{\tenrm
Strangeness chemical potential $\mu_{\rm s}$ versus baryon chemical
potential $\mu_{\rm B}$. The long-dashed line corresponds to $ T= 150$  MeV,
the solid line to $ T = 200$ MeV,
and the dashed line to $T =  300$ MeV.  In Fig.\,\protect\ref{F2}a,
for $\varepsilon\equiv
(\bar s-{s}) /{s} = 0$, the dotted line is the
limiting curve for large $T$ and the horizontal solid line corresponds to
$\mu_{\rm s}=0$ in QGP independent of $T$.
Fig.\,\protect\ref{F2}b, c are the same as Fig.\,\protect\ref{F2}a with
$ \varepsilon \pm 0.1 $ except the dotted
lines corresponding to $T = 1000$ MeV. The horizontal lines give
$\mu_{\rm s}$ in QGP for the corresponding temperatures.
\protect\label{F2}
}
\end{minipage}\hfill
\begin{minipage}[t]{0.475\textwidth}\vspace{-3.3cm}
\centerline{\hspace{0.2cm}\psfig{figure=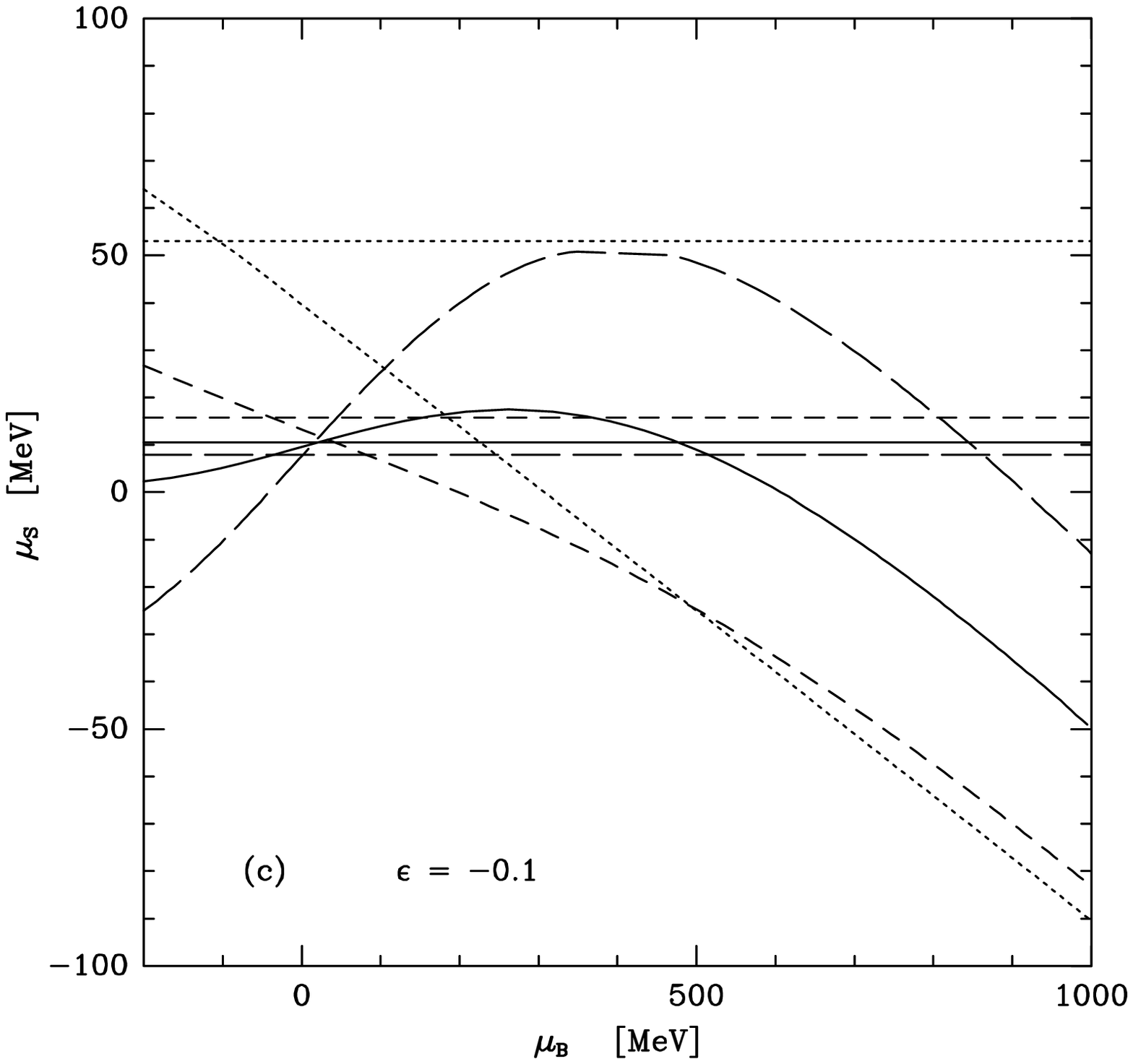,height=11.5cm}}\vspace{-2cm}
\end{minipage}
\end{figure}

For the extensive set of resonances included here this quite peculiar
behavior occurs at $T\simeq210$ MeV for $\varepsilon=0$ and at
$T\simeq200$ MeV for $\varepsilon = - 0.1$. In Figs.\,\ref{F2}a,
\ref{F2}b, \ref{F2}c we  present the constraint between $\mu_{\rm s}$ and
$\mu_{\rm B}$ at  fixed $T=200$ MeV (solid curves), 150 MeV (long-dashed
curves), 300  MeV (short-dashed curves) and 1000 ($\simeq\infty$) MeV
(dotted curves) for $\varepsilon=0$ (Fig.\,\ref{F2}a), $\varepsilon =
0.1$   (Fig.\,\ref{F2}b), and $\varepsilon = -0.1$ (Fig.\,\ref{F2}c), all
 for $\gamma_{\rm s}=0.7$. The choice $\gamma_{\rm s}=1$ would not
change our results significantly. The flat behavior of $\mu_{\rm s}$  as
a function of $\mu_{\rm B}$ for $\varepsilon=0$ and $T\simeq 200$  MeV
makes it hard to distinguish the hadron gas from a QGP (which has
$\mu_{\rm s}\equiv 0$ independent of $\mu_{\rm B}$) in this temperature
range. This was pointed out in Ref.\,\cite{Cley92}, but we think that
Eq.\,(\ref{muB}) and the corresponding Fig.\,\ref{F2}a,  with our
definition of $\mu_{\rm s}$ (see footnote \ref{foot1}), exhibits this
effect more clearly and allows us to understand the  accidental nature of
this behavior.

For $\varepsilon = 0$ (Fig.\,\ref{F2}a) all curves pass through
$\mu_{\rm s}=0$ at $\mu_{\rm B}=0$, and as $T\to \infty$ one
approaches a limiting curve which is rather well represented by the
dotted curve corresponding to 1000 MeV. For $\varepsilon \ne 0$ we  note
that in the QGP phase the strange quark chemical potential does  not
vanish (see horizontal lines in Figures\,\ref{F2}b, \ref{F2}c), and  that
there is a common crossing region of the HG and QGP results at a  finite
value of $\mu_{\rm s}$, close to $\mu_{\rm B}=0$. For $T\to  \infty$  for
a fixed positive (negative) net strangeness fraction  $\varepsilon$,
$\mu_{\rm s}$ continues  to increase (decrease) at  fixed $\mu_{\rm B}$
as a function of $T$, and only the slope as a  function of $\mu_{\rm B}$
approaches a definite limit as $T \to
\infty$.

\pagebreak[5]
$\ $
\vspace{1.3cm}

\noindent

$\ $

\vspace{-1.5cm}

\section{Analysis of the data within the fireball model}
\label{sect3}

In this section we describe, using as an example the experiment with
the most detailed information of strange baryon and anti-baryon
production, WA85 \cite{WA85}, how to extract estimates of the relevant
thermodynamic parameters from particle production data. This procedure
is straightforward if one can focus on the ground state baryons of a
given flavor content, {\it i.e.} if the production and decay of
heavier resonances can be neglected. Such an analysis was presented in
\cite{Raf91,LTR92}, and its essential steps will be shortly repeated
in subsections 3.1 through 3.3. If $T$ is indeed in the region of 210 MeV
as the $m_\bot$-slopes seem to indicate, resonance production and
decays can, however, not be neglected, and therefore such a procedure
is not fully consistent. In subsection \ref{3.4} we therefore redo this
analysis including resonance decays, which is then much more involved.
The main result is that resonance decays mainly affect the extraction
of the strangeness saturation factor $\gamma_{\rm s}$ (inclusion of
resonance decays drives the extracted $\gamma_{\rm s}$ closer to 1,
{\it i.e.} towards full saturation of hadronic gas phase space), while
the striking conclusion that the data point towards a vanishing
strange chemical potential $\mu_{\rm s}\approx 0$ is remarkably stable
against these corrections.

\subsection{Cancellation of the Boltzmann factor}
\label{3.1}

It was argued in \cite{Raf91} that within the thermal model it is quite
easy to extract the fugacities from a given particle ratio {\em at fixed
$m_\bot$ and $y$}, thus eliminating the need to extrapolate the
experimental data to full phase space. Specifically, for a stationary
fireball one finds for the ratio of particles $i$ and $j$, for each of
which the invariant cross section is experimentally known at a certain
rapidity $y\pm dy/2$ in the region $m_\bot \geq m^{\rm cut}_\bot$:
\begin{equation}
   \left. {dN_i/dy \over dN_j/dy} \right\vert_{m_\perp \geq
                                               m_\perp^{\rm cut}}
   = {g_i\, \gamma_i\, \lambda_i \over g_j\ \gamma_j\, \lambda_j}\,
{\int_{m^{\rm cut}_\perp}^\infty m_\perp\, \cosh y\,
      \exp(-m_\perp\cosh y/T)\, dm_\perp^2 \over
      \int_{m^{\rm cut}_\perp}^\infty m_\perp\, \cosh y\,
      \exp(-m_\perp\cosh y/T)\, dm_\perp^2}\ .
 \label{NiNj}
 \end{equation}
Here we used the identity $E=m_\perp\cosh y$ for the particle energy
(with $y$ measured relative to the center of mass of the collision
fireball) to rewrite the transverse mass integral over the Boltzmann
distribution $E\, \exp(-E/T)$ (the factor $E$ results from  starting
with the Lorentz-invariant cross section $E\, d^3N/d^3p$). If the
fireball  is not stationary, but expands in the longitudinal direction
in a  boost-invariant way, $dN/dy$ is independent of $y$, and the
factors  $E\, \exp(-E/T)$ under the integrals on the r.~h.~s. of
Eq.~(\ref{NiNj}) have to be replaced by a Bessel function \cite{SH92}:
\begin{equation}
   {dN_i^{\rm thermal}\over dy\, dm_\perp^2}
   = g_i\, \gamma_i\, \lambda_i \, m_\perp\,
      K_1\left({m_\perp\over T}\right)\, ,
 \label{boost}
 \end{equation}
where we neglected a common normalization constant containing the
fireball volume which drops out from the ratio\footnote{
The same result is obtained if the spectrum in Eq.~(\ref{NiNj}) is
integrated over rapidity, {\it i.e.} if one wants to describe with a
stationary fireball data obtained over a large rapidity window.}.
The important observation is that obviously in both cases the
Boltzmann factors (and with them all dependence on the particle rest
masses) drop out from the ratio, due to the common lower cutoff on
$m_\perp$ (which, of course, must be larger than either rest mass).
The same remains true if $m_\perp$ is not integrated over, but the
ratio is evaluated at fixed $m_\perp$. Since the degeneracy factors
$g_i$ are known, this immediately yields the ratio of the generalized
fugacities $\gamma_i\lambda_i$ (see Eq.\,(\ref{gammai})).

\subsection{Extraction of $\mu_{\rm s}$ and $\mu_{\rm q}$
from anti-baryon data}
\label{3.2}

Thus, comparing the {\it spectra\/} of particles with those of their
anti-particles within overlapping regions of $m_\bot$, the Boltzmann
and {\em all} other statistical factors cancel from their relative
normalization, which is thus a function of the fugacities only. [In
subsection\ref{3.4} we show that this statement, although no longer
strictly true, still remains approximately valid when feed-down by
resonance decays is included.] For the currently available two such
ratios one has specifically
 \begin{eqnarray}
  R_\Xi &&\hspace{-0.6cm}=  {{\overline{\Xi^-}}\over {\Xi^-}} =
   {{\lambda_{\rm d}^{-1} \lambda_{\rm s}^{-2}} \over
    {\lambda_{\rm d} \lambda_{\rm s}^2}} \ ,
 \label{ratioL}\\
  R_\Lambda&&\hspace{-0.6cm}= {\overline{\Lambda}\over \Lambda} =
   {{\lambda_{\rm d}^{-1} \lambda_{\rm u}^{-1}
                          \lambda_{\rm s}^{-1}} \over
    {\lambda_{\rm d} \lambda_{\rm u} \lambda_{\rm s}}} \ .
 \label{ratio}
 \end{eqnarray}

These ratios can easily be related to each other, in a way which shows
explicitly the respective isospin asymmetry factors and strangeness
fugacity dependence. Eqs.\,(\ref{ratioL}, \ref{ratio}) imply:
 \begin{eqnarray}
   R_\Lambda R_\Xi^{-2} &&\hspace{-0.6cm}=  e^{6\mu_{\rm s}/T}\cdot
e^{2\delta\mu/T}\ ,   \label{R1}\\
   R_\Xi R_\Lambda^{-2}&&\hspace{-0.6cm}= e^{6\mu_{\rm q}/T} \cdot
e^{-\delta\mu/T}\ .   \label{R2}
 \end{eqnarray}
Eqs.\,(\ref{R1}, \ref{R2}) are generally valid, irrespective of the
state of the system (HG or QGP), as long as the momentum spectra of
the radiated particles are thermal with a common temperature. We see
that once the left hand side is known experimentally, it determines
rather accurately the values of $\mu_{\rm q},\mu_{\rm s}$ which enter
on the right hand side with a dominating factor 6, while the (small)
isospin asymmetry $\delta\mu$ plays only a minor role. This explains
how, by applying these identities to the WA85 data \cite{WA85}, it has
been possible to determine the chemical potentials with considerable
precision in spite of the still pretty large experimental errors on
the measured values of $R_\Lambda$, $R_\Xi$.

Turning to the pertinent experimental results we first recall the
recently released data on the rapidity dependence of
$\overline{\Lambda}$ and $\Lambda$ production rates measured by the
experiment NA36 \cite{NA36}. There is a clear indication that the
production of these particles in nuclear S--Pb collisions is
concentrated in the central rapidity domain --- it does not follow in
its behavior the FRITIOF simulations which, on the other hand, are
consistent with the production rates from p--Pb collisions. This
supports the notion of a central fireball reaction picture for S--W
collisions studied by WA85 \cite{WA85}. The most striking result of
the latter experiment is the fact that the abundance of charged
anti-cascades ($\bar s\bar s\bar d$) is unusually enhanced, in particular
when compared to the abundance of cascades ($ssd$) and anti-lambdas
($\bar s\bar u \bar d$).  We now interpret the data in terms of the
local equilibrium model, thus fixing the values of $\mu_{\rm s}$ and
$\mu_{\rm q}$.

The $\overline{\Xi^-}/ \Xi^-$ ratio has been reported as:
 \begin{eqnarray}
   R_\Xi = 0.39\pm 0.07\  \quad \mbox{ for }
   y \in (2.3,3.0) \mbox{ and } m_\bot>1.72\ \mbox{GeV}.
 \label{cascade}
 \end{eqnarray}
Note that, in p--W reactions in the same $(p_\bot,y)$ region, a smaller
value for the $R_\Xi$ ratio, namely $0.27\pm 0.06$, is found. The  ${\bar
\Lambda}/\Lambda $ ratio is:
 \begin{eqnarray}
   R_\Lambda = 0.13\pm 0.03\  \quad \mbox{
   for } y \in (2.3,3.0) \mbox{ and } m_\bot>1.72\ \mbox{GeV}.
 \label{lambda}
 \end{eqnarray}
In Eq.\,(\ref{lambda}), corrections were applied to eliminate hyperons
originating from cascade decays, but not those originating from decays of
$\Omega \to \Lambda + \overline{K}$ or $\overline{\Omega} \to
\overline{\Lambda} + K$ which are of little significance for the high
$m_\perp$ considered here. The ratio $R_\Lambda$ for S--W collisions is
slightly smaller than for p--W collisions in the same kinematic range.

{}From these two results, together with Eqs.\,(\ref{R1}, \ref{R2},
\ref{deltamu}) and the value $R_{\rm f}=1.08$ mentioned after
Eq.~(\ref{eq4}), we obtain the following values of the chemical
potentials for S--W central collisions at 200 GeV A:
 \begin{eqnarray}
   {\mu_{\rm q}\over T} &&\hspace{-0.6cm}=  {\ln R_\Xi/R_\Lambda^2 \over
5.92}                          = 0.53 \pm 0.1\, ,
  \label{muq}\\
   {\delta\mu\over T}   &&\hspace{-0.6cm}=  {\mu_{\rm q}\over T}(R_{\rm
f}-1)                          = 0.042 \pm 0.008\, ,
  \label{dmu}\\
   {\mu_{\rm s}\over T} &&\hspace{-0.6cm}=  {\ln R_\Lambda/R_\Xi^2\
-0.084\over 6}                          = -0.018 \pm 0.05\, .
  \label{mus}
 \end{eqnarray}
A study of the temperature observed in these collisions was recently made
\cite{Raf92}. As noted in \cite{LTR92} and discussed below in section~4,
the possibility of a HG interpretation of these data only poses itself
if the transverse mass slope of the produced particles is entirely due
to the thermal motion. Under this assumption the temperature is $T =  210
\pm 10$ MeV, and therefore we have:
 \begin{equation}
  \mu_{\rm B} = 335 \pm 25 \mbox{ MeV}\ ,\quad
  \delta\mu = 9 \pm 2 \mbox{ MeV}\ ,\quad
  \mu_{\rm s} = -3.8 \pm 10 \mbox{ MeV}\ .
  \label{chempot}
 \end{equation}
The last result translates into the value $\lambda_{\rm s} = 0.98 \pm
0.05$ for the strange quark fugacity. Much of what we do later in this
paper will rest on this stunning result that the strange chemical
potential is very small and perfectly compatible with zero. In our
calculations below we will thus consider in particular a HG
constrained to this range of values (as well as to near strangeness
neutrality).

\subsection{Determination of $\gamma_{\rm s}$}
\label{3.3}

A complete cancellation of the fugacity factors occurs when we form
the product of the abundances of baryons and anti-baryons. We further
can take advantage of the cancellation of the Boltzmann factors by
comparing this product for two different particle kinds, {\it e.g.} we
consider:
 \begin{equation}
   \Gamma_{\rm s} \equiv \left. {\Xi^-\over\Lambda}
   \cdot {\overline{\Xi^-}\over \overline{\Lambda}
   }\right\vert_{m_\perp>m_\perp^{\rm cut}} \, .
  \label{gam1}
 \end{equation}
If the phase space of strangeness, like that of the light flavors, were
fully saturated, the fireball model would imply $\Gamma_{\rm s}=1$.
However, any deviation from absolute chemical equilibrium as expressed
by the factor $\gamma_{\rm s}$ will change the value of $\Gamma_{\rm s}$.
Ignoring as before any feed-down effects from higher resonances we  find
\begin{equation}
    \Gamma_{\rm s}=\gamma_{\rm s}^2\, .
 \label{gam2}
 \end{equation}
In principle the measurement of $\gamma_s$ can be done with other
particle ratios; in the absence of resonance feed-down we have
 \begin{equation}
   \gamma_{\rm s}^2 = \left.
     {\Lambda\over p} \cdot {\overline{\Lambda}\over \overline p}
              \right\vert_{m_\perp>m_\perp^{\rm cut}}
                    = \left.
     {\Xi^-\over\Lambda} \cdot {\overline{\Xi^-} \over
       \overline{\Lambda}}
       \right\vert_{m_\perp>m_\perp^{\rm cut}}
                    = \left.
     {\Omega^-\over 2\Xi^-} \cdot {\overline{\Omega^-} \over
                                            2\overline{\Xi^-}}
                      \right\vert_{m_\perp>m_\perp^{\rm cut}} \, ,
\label{gam3}
 \end{equation}
where in the last relation the factors 2 in the denominator correct
for the spin-3/2 nature of the $\Omega$. As we will see in subsection 3.5,
Eq.~(\ref{gam4}), these double ratios are considerably affected by
resonance decays.

We note that in the kinematic domain of Eqs.\,(\ref{cascade},
\ref{lambda}) the experimental results reported by the WA85
collaboration are:
 \begin{equation}
   \frac{\overline{\Xi^-}}{\overline{\Lambda}+\overline{\Sigma^0}} =
0.6 \pm 0.2\, , \quad
   \frac{\Xi^-}{\Lambda+\Sigma^0} = 0.20 \pm 0.04\, .
 \label{newa}
 \end{equation}
If the mass difference between $\Lambda$ and $\Sigma^0$ is neglected,
this implies in the framework of the thermal model that an equal
number of $\Lambda$'s and $\Sigma^0$'s are produced, such that
 \begin{equation}
   \frac{\overline{\Xi^-}}{\overline \Lambda}
   = 1.2 \pm 0.4\, , \hspace{1.15cm}
   \frac{\Xi^-}{\Lambda} = 0.40 \pm 0.08\, .
 \label{new1}
 \end{equation}
The fact that the more massive and stranger anti-cascade exceeds at
fixed $m_\bot$ the abundance of the anti-lambda is most striking.
These results are inexplicable in terms of simple cascade models for
the heavy-ion collision \cite{Csernai}. The relative yield of
$\overline{\Xi^-}$ appears to be 5 times greater than seen in the
$p$--$p$ ISR experiment \cite{ISR} and all other values reported in
the literature \cite{WA85}.

Combining the experimental result Eq.\,(\ref{new1}) with Eqs.
(\ref{gam1}, \ref{gam2}), we find the value
 \begin{equation}
  \gamma_{\rm s}=0.7 \pm 0.1\, .
 \label{gams}
 \end{equation}

\subsection{Temperature, transverse mass spectrum}
\label{3.4}
The data from the WA85 collaboration \cite{WA85} correspond to an
apparent temperature (from the slope of the $m_\perp$-spectra) of $T_{\rm
apparent} = 210\pm 10$ MeV \cite{Raf92}. In our numerical  studies we
therefore concentrated on two extreme cases:

\begin{itemize}

\item[i)] $T=210$ MeV, $\beta_\bot=0$ (no transverse flow), which we call
the ``thermal" scenario,

\item[ii)] $T=150$ MeV, $\beta_\bot=0.32$, which we call the ``flow"
scenario for which the freeze-out temperature has been estimated
following the results of \cite{SH92}.

\end{itemize}

If the slope of the experimental spectra is not entirely due to
thermal motion, but contains a flow component due to collective
expansion of the emitting source \cite{SH92}, then ``flow spectra"
({\it i.e.}  boosted Boltzmann spectra) have to be used in the numerator
and  denominator of Eq.~(\ref{NiNj}). If the flow is azimuthally
symmetric  and boost-invariant in the longitudinal direction, the thermal
spectrum (\ref{boost}) has to be replaced by (again neglecting a
common normalization constant)
 \begin{equation}
   {dN_i^{\rm flow}\over dy\, dm_\perp^2}
   = g_i\, \gamma_i\, \lambda_i \, m_\perp\,
      K_1\left({m_\perp\cosh\rho \over T}\right)\,
      I_0\left({p_\perp\sinh\rho \over T}\right)\, ,
 \label{flow}
 \end{equation}
where $K_1$ and $I_0$ are the modified Bessel functions and
 \begin{equation}
  \rho = \tanh^{-1} \beta_f
 \label{rho}
 \end{equation}
is the transverse flow rapidity. This spectrum has an asymptotic
slope corresponding to an apparent ``blue-shifted" temperature
 \begin{equation}
   T_{\rm apparent} = T \sqrt{{1+\beta_f}
                        \over {1-\beta_f}}\, .
 \label{Tapp}
 \end{equation}
The same result as in Eq.\,(\ref{flow}) is obtained if the spectrum
from a spherically expanding fireball is integrated over rapidity.
Since some longitudinal flow is likely to be present in S--W
collisions, and also the kinematic window of the WA85 experiment is
nearly one unit of rapidity, we think that the form (\ref{flow}), in
which integration over a large rapidity window is implied, is most
appropriate for an analysis of the experimental spectra. (The ``large
window" combines the one unit of $y$-acceptance in WA85 with about 1.5
units of flow expected \cite{SH92} for this system.)

Obviously, through the separate dependence on $m_\perp$ and $p_\perp$
Eq.~(\ref{flow}) depends on the particle rest mass explicitly. This
spoils the cancellation of the spectral shape from the
$m_\perp$-integrated particle ratios as well as their independence
from the $m_\perp$-cut, and thus also renders the extraction of the
fugacities, etc. more complicated.

In order to assess the importance of these effects we devote next
subsection to a more complex reanalysis of the experimentally measured
strange baryon and anti-baryon ratios which takes all these complications
into account. Anticipating this development we show in Fig.\,\ref{F5new}
the $m_\perp$-spectra of the various particle species, for both the
``thermal" scenario (Figs.\,\ref{F5new}a, b) and the ``flow" scenario
(Figs.\,\ref{F5new}c, d). Dashed lines indicate the direct thermal
contribution to the spectra, while the solid lines include all resonance
decay contributions (see following subsection). These results allow to
directly assess the importance of resonance decays on the
$m_\perp$-slopes of the observed particles and on the particle ratios.
For the $\Omega$ there are no resonance contributions, due to the
imposed mass cutoff in the resonance spectrum. While the overall
normalization of the spectra is arbitrary, their relative
normalization is according to the fugacities given in Table~1.

The presence of flow affects the spectra mostly at low $m_\perp$ near
the mass threshold \cite{SH92}, where it leads to a flattening, {\it
i.e.}  to a higher apparent temperature. The effect is proportional to
the  particle mass (since for small $p_\perp$ the effect of flow can be
estimated by $p_{\rm flow} \simeq m_0\, v_{\rm flow}$) and thus most
clearly seen in the $\Xi^- $ and $\Omega$ spectra.

\subsection{Resonance decays}
\label{3.5}

As already mentioned, Eq.\,(\ref{NiNj}) is only correct if feed-down
to the observed particles by strong decays of higher resonances is
neglected. At temperatures of order 200 MeV these resonance decays can
no longer be totally ignored. This has two annoying consequences:
first, since the decay products have a steeper $m_\perp$-spectrum than
the original resonances \cite{sollfrank}, the various contributions to
numerator and denominator in (\ref{NiNj}) have different slopes, and
the integrals over the Boltzmann factors no longer cancel exactly.
Second, since different sets of resonances contribute to $i$ and $j$,
the ratio is modified by an a priori unknown factor which turns out to
be hard to calculate and in general {\em does} depend on the
experimental $m_\perp$-cutoff as well as on the temperature.

Since the decay spectra drop more steeply as a function of $m_\perp$
than the thermal ones, the role of resonance feed-down can in
principle be suppressed by going to sufficiently large $m_\perp$. For
pions and kaons, for example, the thermal contribution dominates for
$m_\perp >$ 1--1.5 GeV, in spite of the huge contribution of decay
pions to the total yield \cite{sollfrank}; the latter all end up at
lower $m_\perp$. For baryons  due  to their larger masses, the
difference in slope between the thermal and decay baryons is less
drastic, and one has to go to still larger $m_\perp$ before the
resonance decay contributions are sufficiently suppressed.

Including resonance decays, the numerator and denominator of
Eq.~(\ref{NiNj}) take the form
 \begin{equation}
   \left. {dN_i \over dy} \right\vert_{m_\perp \geq
                                               m_\perp^{\rm cut}}
   =  \int_{m^{\rm cut}_\perp}^\infty dm_\perp^2
      \left\{ {dN_i^{\rm thermal/flow}(T) \over dy\, dm_\perp^2} +
    \sum_R b_{R\to i} {dN_i^R(T) \over dy\, dm_\perp^2} \right\}\, ,
\label{reso}
 \end{equation}
with the direct thermal or flow contribution given by (\ref{boost}) or
(\ref{flow}), respectively, and the contribution from decays of
resonances $R\to i+2+\dots+n$ (with branching ratio $b_{R\to i}$ into
the observed channel $i$) calculated according to \cite{sollfrank}
\begin{eqnarray}
   \label{eq:reso}
    {{dN_i^R}\over{dy\,dm_\perp^2}} =
    \int_{s_-}^{s_+} ds \,g_n(s)
    \!\!\!&&\!\!\! \int_{Y_-}^{Y_+} dY \int_{M_\perp^-}^{M_\perp^+}
dM_\perp^2
  \\ \nonumber
    && {{M}\over{\sqrt{P_\perp^2 p_\perp^2
                 -[ME^* - M_\perp m_\perp \cosh(Y-y)]^2}}}
       \left( {{dN_R}\over{dY \,dM_\perp^2}} \right)\, .
 \end{eqnarray}
Capital letters indicate variables associated with the resonance
$R$, $\sqrt{s}$ is the invariant mass of the unobserved decay
products $2,\dots,n$, and the kinematic limits are given by
 \begin{eqnarray}
   s_- \!\!&&\hspace{-0.6cm}= \!\! \left( \sum_{k=2}^n m_k \right)^2 ,
\qquad    s_+ = (M-m_i)^2\, ,
  \label{spm}\\
   Y_\pm \!\!&&\hspace{-0.6cm}= \!\! y \pm \sinh^{-1}
{{p^*}\over{m_\perp}}\, ,   \label{eq:ylim}\\
   M_\perp^\pm \!\!&&\hspace{-0.6cm}= \!\! M
   {E^*m_\perp\cosh(Y-y) \pm p_\perp
    \sqrt{{p^*}^2 - m_\perp^2 \sinh^2(Y-y)} \over
    m_\perp^2 \sinh^2(Y-y) + m_i^2 } \, ,
  \label{mlim}\\
   E^* \!\!&&\hspace{-0.6cm}= \!\! {1\over{2M}}( M^2 + m_i^2 - s )\, ,
\qquad p^* = \sqrt{{E^*}^2-m_i^2} \, .
 \end{eqnarray}
In (\ref{eq:reso}) $g_n(s)$ is the decay phase space for the $n$-body
decay $R\to  i+2+\dots+n$, and for the dominant 2-body decay (assuming
isotropic  decay in the resonance rest frame) it is given by
 \begin{eqnarray}
   g_2(s) &&\hspace{-0.6cm}=  {1\over{4\pi p^*}}\, \delta(s-M^2)\, ,
\label{2body}
 \end{eqnarray}
for $n>2$ see Ref. \cite{sollfrank}.

The resonance spectrum $dN_R/dY\, dM_\perp^2$ entering the r.~h.~s. of
Eq.~(\ref{eq:reso}) is in general itself a sum of a thermal or flow
spectrum like (\ref{boost}) or (\ref{flow}) and of decay spectra from
still higher lying resonances; for example, most decays of high-lying
nucleon resonances into protons or anti-protons proceed through the
$\Delta(1232)$ or its anti-particle ({\it e.g.} $N(1440)\to \Delta(1232)
+   \pi \to p + 2\pi$). This has been taken into account in our
calculations. Extracting the degeneracy factors and fugacities of the
decaying resonances, we write shortly
 \begin{equation}
  N^R_i \equiv \gamma_R \lambda_R \tilde N^R_i
        \equiv \gamma_R \lambda_R \int_{m_\perp^{\rm cut}}^\infty
               dm_\perp^2 \sum_R  g_R\,b_{R\to i}
               {d\tilde N^R_i \over dy\, dm_\perp^2} \, ,
 \label{ntilde}
 \end{equation}
with the sum now including only resonances with identical quantum
numbers; if these quantum numbers agree with those of species $i$, the
sum is meant to also include the thermal contribution.

Between particles and anti-particles we have the relation
 \begin{equation}
   N_{\bar i}^{\bar R} =  \gamma_R\, \lambda_R^{-1}\, {\tilde N}_i^R
                 = \lambda_R^{-2}\, N_i^R \, .
 \label{ntildea}
 \end{equation}
The three independent particle ratios measured by WA85 \cite{WA85} are
thus given  by
 \begin{eqnarray}
   R_\Xi &&\hspace{-0.6cm}=  \left. {\overline{\Xi^-} \over \Xi^-}
   \right\vert_{m_\perp\geq m_\perp^{\rm cut}}
         = {\gamma_{\rm s}^2 \lambda_{\rm q}^{-1} \lambda_{\rm s}^{-2}
        {\tilde N}_\Xi^{\Xi^*} +
            \gamma_{\rm s}^3 \lambda_{\rm s}^{-3}
            {\tilde N}_\Xi^{\Omega^*} \over
            \gamma_{\rm s}^2 \lambda_{\rm q} \lambda_{\rm s}^2
            {\tilde N}_\Xi^{\Xi^*} +
            \gamma_{\rm s}^3 \lambda_{\rm s}^3
            {\tilde N}_\Xi^{\Omega^*} }\ ,
 \label{rxi}\\
   R_\Lambda &&\hspace{-0.6cm}=  \left. \hphantom{^-}{\overline{\Lambda}
           \over \Lambda}
             \right\vert_{m_\perp\geq m_\perp^{\rm cut}}
         = {\gamma_{\rm s} \lambda_{\rm q}^{-2} \lambda_{\rm s}^{-1}
      {\tilde N}_\Lambda^{Y^*} +
            \gamma_{\rm s}^2 \lambda_{\rm q}^{-1} \lambda_{\rm s}^{-2}
        {\tilde N}_\Lambda^{\Xi^*} +
            \gamma_{\rm s}^3 \lambda_{\rm s}^{-3}
            {\tilde N}_\Lambda^{\Omega^*} \over
            \gamma_{\rm s} \lambda_{\rm q}^2 \lambda_{\rm s}
            {\tilde N}_\Lambda^{Y^*} +
            \gamma_{\rm s}^2 \lambda_{\rm q} \lambda_{\rm s}^2
            {\tilde N}_\Lambda^{\Xi^*} +
            \gamma_{\rm s}^3 \lambda_{\rm s}^3
            {\tilde N}_\Lambda^{\Omega^*} }\, ,
 \label{rlam}\\
   R_{\rm s} &&\hspace{-0.6cm}=  \left. {\Xi^- \over \Lambda}
           \right\vert_{m_\perp\geq m_\perp^{\rm cut}}
         = {\gamma_{\rm s}^2 \lambda_{\rm q} \lambda_{\rm s}^2
            {\tilde N}_\Xi^{\Xi^*} +
            \gamma_{\rm s}^3 \lambda_{\rm s}^3
            {\tilde N}_\Xi^{\Omega^*}  \over
            \gamma_{\rm s} \lambda_{\rm q}^2 \lambda_{\rm s}
            {\tilde N}_\Lambda^{Y^*} +
            \gamma_{\rm s}^2 \lambda_{\rm q} \lambda_{\rm s}^2
            {\tilde N}_\Lambda^{\Xi^*} +
            \gamma_{\rm s}^3 \lambda_{\rm s}^3
            {\tilde N}_\Lambda^{\Omega^*} }    \, .
 \label{rs}
 \end{eqnarray}
$\tilde N_\Lambda^{Y^*}$ contains also (in fact as its most important
contribution) the electromagnetic decay $\Sigma^0 \to \Lambda +
\gamma$, and $\tilde N_\Lambda^{\Omega^*}$ includes the weak decay
$\Omega \to YK$ (since $\Omega$'s have not been reconstructed so far
by WA85). If the right hand sides are evaluated using
Eqs.~(\ref{reso}, \ref{eq:reso}, \ref{ntilde}) with $m_\perp^{\rm
cut}= 1.72$ GeV, for the left hand sides the experimental values
(\ref{cascade}, \ref{lambda}, \ref{newa}) can be inserted, and
Eqs.~(\ref{rxi}, \ref{rlam}, \ref{rs}) can be solved for $\gamma_{\rm
s}$, $\lambda_{\rm s}$, and $\lambda_{\rm q}$. Note that we do not
distinguish here between $\lambda_{\rm u}$ and $\lambda_{\rm d}$ since
this would require the calculation of a separate decay cascade for
each member of an isospin multiplet.  The smallness of $\delta\mu$ in
(\ref{dmu}) justifies this approximation.

\begin{figure}[ht]
\vspace{ -4cm}
\centerline{\psfig{figure=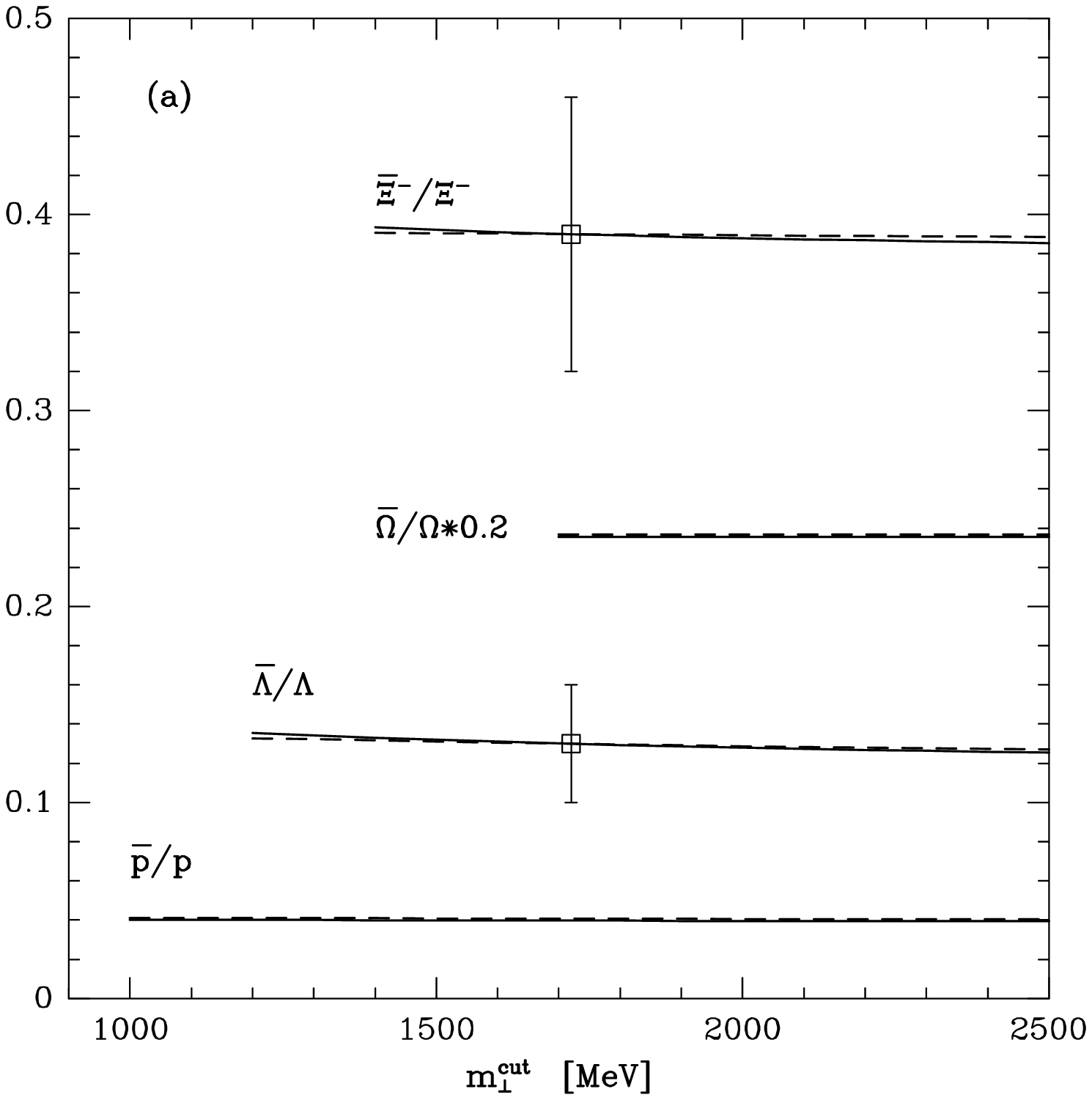,height=11.5cm}
\psfig{figure=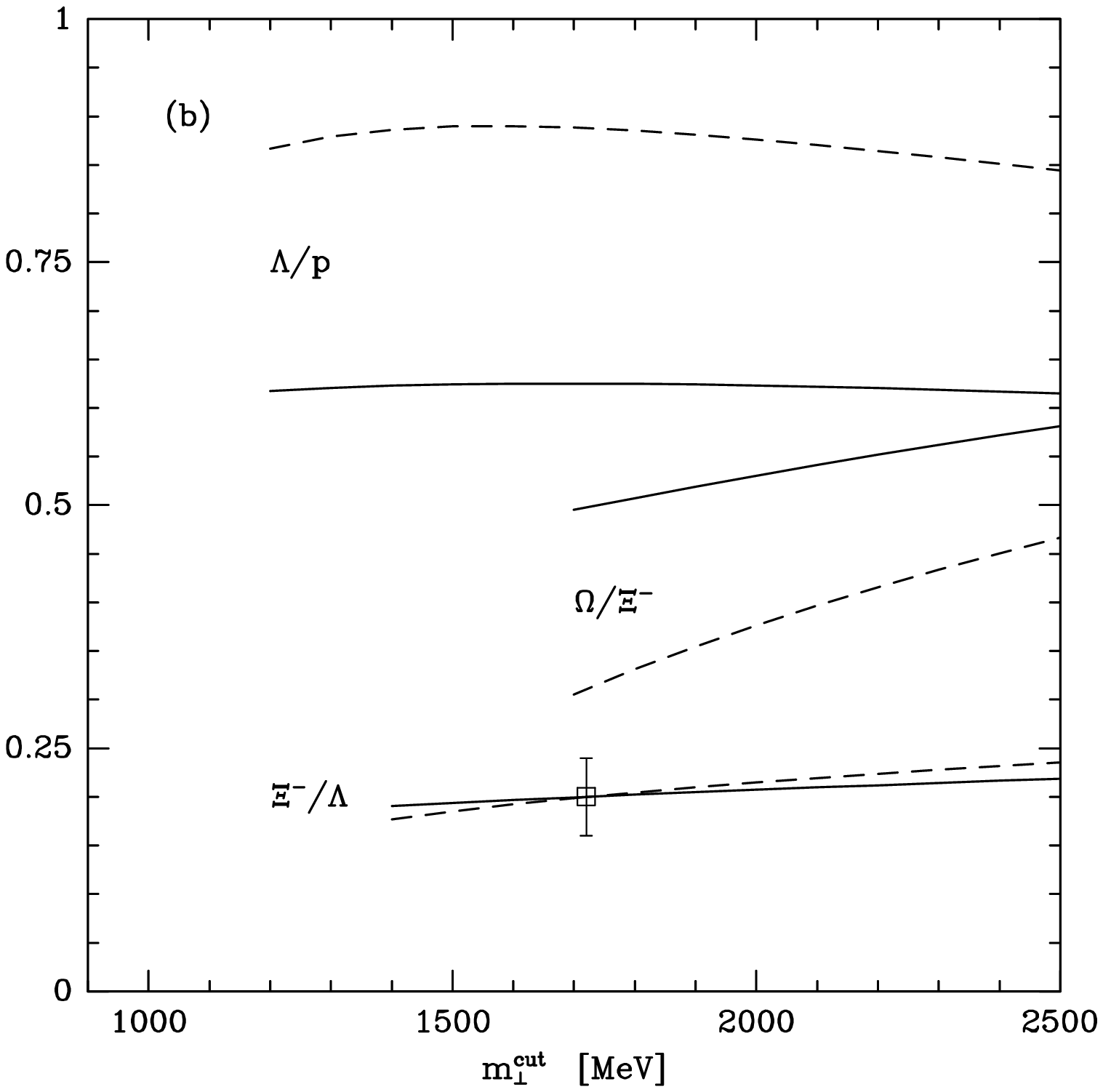,height=11.5cm}}
\vspace{ -3.5cm}
\centerline{\psfig{figure=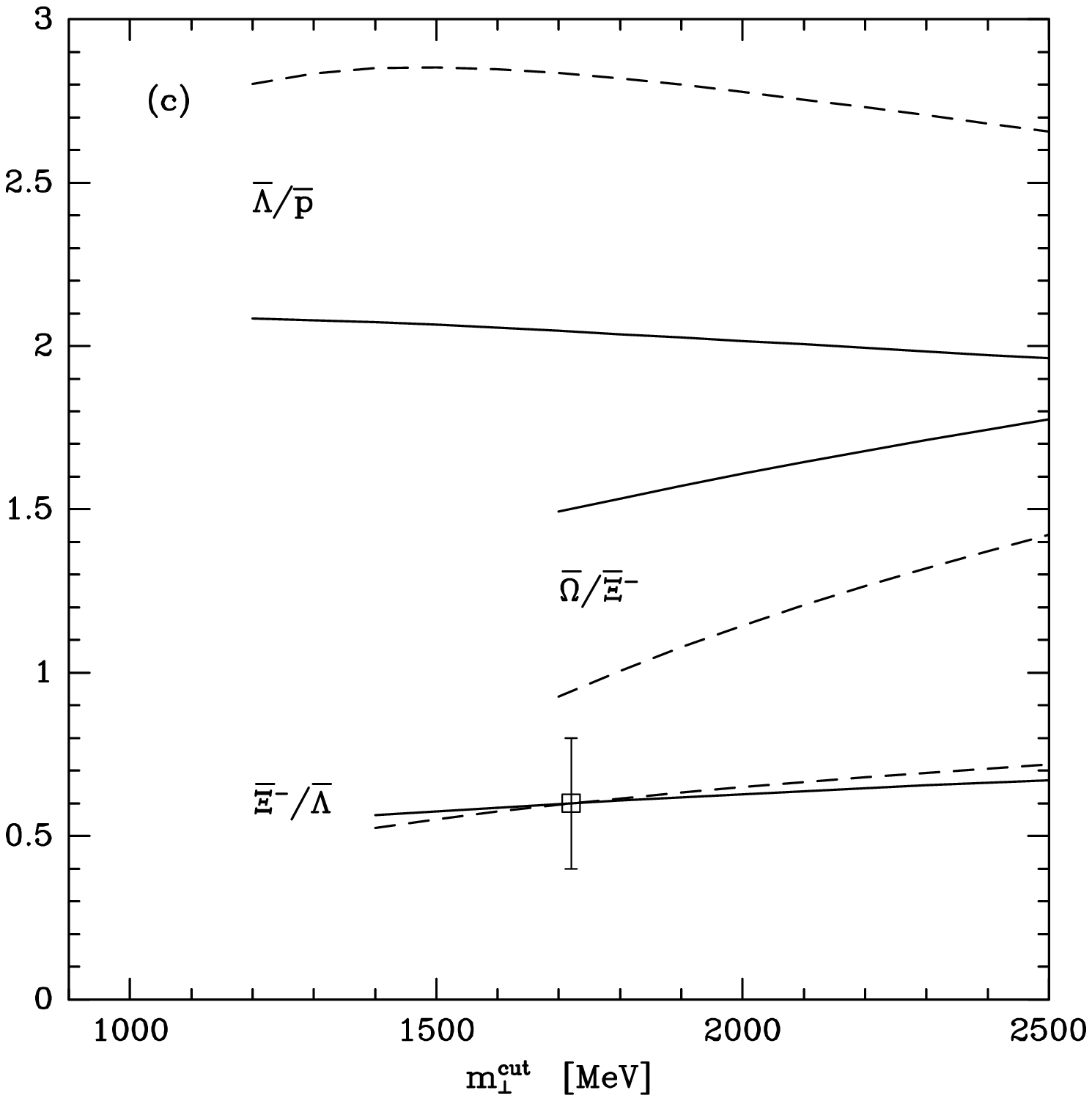,height=11.5cm}
\psfig{figure=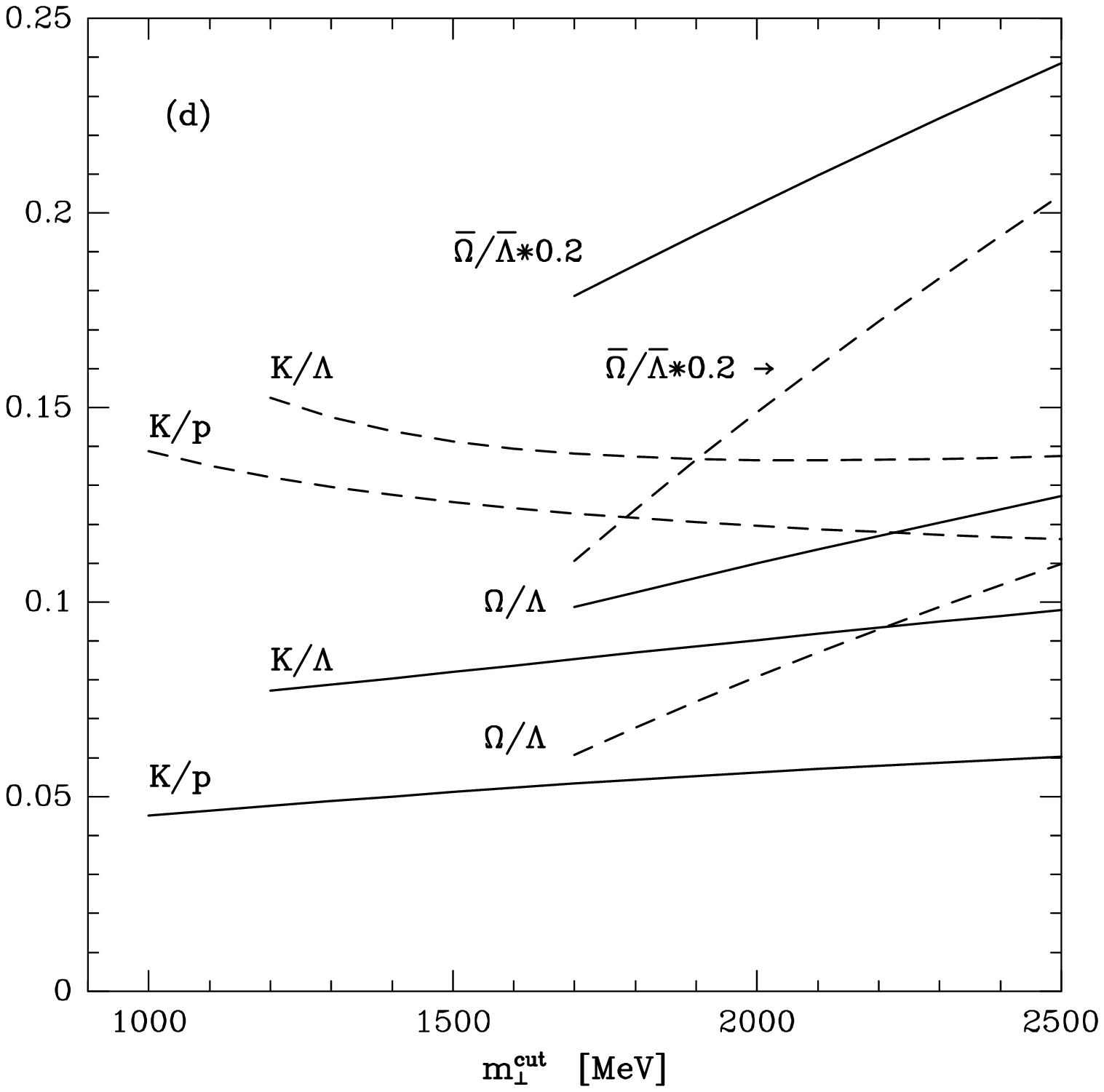,height=11.5cm}}
\vspace{ -2cm}

\caption{\tenrm
Various $m_\perp$-integrated particle ratios as a function of the
low-$m_\perp^{cut}$. Solid lines are for the ``thermal" scenario, dashed
lines are for the ``flow" scenario, with parameters as given in Table
1.
\label{F6new}
}
\end{figure}

The results of our analysis are given in Tables 1 and 2 and Figures
\ref{F5new} and \ref{F6new}. Comparing the results for the thermal
parameters in Table 1 with Eqs.~(\ref{muq}, \ref{mus}, \ref{gams}) which
contained only an estimate of the radiative $\Sigma^0$ decay but no
contributions from  higher lying resonances, we can draw the following
conclusions:

\begin{itemize}

\item[1.] The extraction of $\lambda_{\rm q}$ and $\lambda_{\rm s}$
according to Eqs.~(\ref{ratioL}, \ref{ratio}) is only very weakly
affected by resonance decays and by the origin of the slope of the
$m_\perp$-spectrum (thermal or flow). (The absolute value of the
associated chemical potentials does, of course, depend on the choice
of the freeze-out temperature.) The reason is that in Eqs. (\ref{rxi})
and (\ref{rlam}) in each case the first term in the sums occurring in
the numerator and denominator completely dominates. (For $\Xi^- $'s
the effects from $\Omega$-decays are small, and for $\Lambda$'s the
strong $\Sigma^0$ contribution overwhelms the effect from $\Xi^*$
decays.) Taking the results for the ``thermal" scenario from Table 1
and evaluating with them the $\overline{\Xi^- } / \Xi^- $,
$\overline{\Lambda} / \Lambda$, and $\Xi^- /\Lambda$ ratios excluding
the contributions from higher resonances changes them to 0.383, 0.121,
and 0.262, respectively (instead of the input values 0.39, 0.13, and
0.20, respectively). Leaving out also the $\Sigma^0\to\Lambda\gamma$
contribution changes only the $\Xi^- /\Lambda$ ratio which now becomes
0.430. Obviously this ratio is the one which is most strongly affected
by the resonance decays.

\item[2.] Since the $\Xi^- /\Lambda$ ratio is the crucial ingredient for
the determination of the strangeness saturation factor $\gamma_{\rm  s}$,
the effects from resonance decays and flow can be clearly seen in  this
number. From Eq.~(\ref{new1}) we had extracted $\gamma_{\rm s}=0.7$;
improving the rough factor 2 estimate for the $\Sigma^0$ contribution
by its correct thermal weight at $T=210$ MeV would have lowered
$\gamma_{\rm s}$ to 0.57. The decay of higher-lying resonances raises
the value of $\gamma_{\rm s}$ up to 0.76 at $T=210$ MeV. If due the
presence of flow the $m_\perp$-slope does not give~the~true

\pagebreak[5]
$\ $

\vspace{2.5cm}
\begin{table}[ht]
\vspace{-4.cm}
\centerline{
 \vbox{\tabskip=0pt \offinterlineskip
 \def\tablerule{\noalign{\hrule}}
 \halign to12cm {\strut#&
  \vrule#\tabskip=0em plus 1em& \hfil#& \vrule#& \hfil#\hfil&
  \vrule#& \hfil#& \vrule# \tabskip=0pt\cr
\tablerule
   && && && & \cr
   &&\hfil && \hfil (a) pure ``thermal" $\!\!\!\phantom{\Big\vert}$
      \hfil && \hfil (b) ``thermal \& flow" \hfil & \cr
         &&\hfil && \hfil $T=210$ MeV, $\beta_f=0$ \hfil &
       & \hfil $T=150$ MeV, $\beta_f=0.32$ \hfil &\cr            && &&
&& & \cr
\tablerule
           && && && &\cr
           &&\hfil $\lambda_{\rm s}$ \hfil &&
             \hfil \phantom{--}0.973 \hfil && \hfil \phantom{--}0.972
         \hfil &\cr
           && \hfil $\mu_{\rm s}/T$ \hfil &
            & \hfil --0.027 \hfil & & \hfil --0.028 \hfil &\cr
           && \hfil $\mu_{\rm s}$ (MeV) \hfil &
            & \hfil --5.7\phantom{00} \hfil & & \hfil --4.2\phantom{00}
          \hfil &\cr
           && && && & \cr
\tablerule
           && && && & \cr
           && \hfil $\lambda_{\rm q}$ \hfil &
            & \hfil \phantom{11}1.718 \hfil & & \hfil \phantom{0}1.710
           \hfil &\cr
           && \hfil $\mu_{\rm q}/T$ \hfil &
            & \hfil \phantom{11}0.54\phantom{0} \hfil & &
              \hfil \phantom{0}0.54\phantom{0}
              \hfil &\cr
           && \hfil $\mu_{\rm q}$ (MeV) \hfil &
            & \hfil 113.6\phantom{00} \hfil & & \hfil 80.5\phantom{00}
           \hfil &\cr
           && && && & \cr
\tablerule
           && && && &\cr
           && \hfil $\gamma_{\rm s}$ \hfil &
            & \hfil 0.756 \hfil & & \hfil 0.902 \hfil &\cr
           && && && &\cr
\tablerule
}
}
}
\vspace{-0.2cm}
\caption{
\tenrm Thermal fireball parameters extracted from the
WA85 data \protect\cite{WA85} on strange baryon and anti-baryon production, for
two different interpretations of the measured $m_\bot$-slope.
Resonance decays were included. For details see text.}

$\ $
\vfill
\centerline{
 \vbox{\tabskip=0pt \offinterlineskip
 \def\tablerule{\noalign{\hrule}}
 \halign to16cm {\strut#&
  \vrule#\tabskip=0em plus 1em& \hfil#& \vrule#& \hfil#\hfil&
  \vrule#& \hfil#\hfil& \vrule#& \hfil#\hfil&
  \vrule#& \hfil#& \vrule# \tabskip=0pt\cr
\tablerule
           && \hfil \lower 8pt\hbox{$N_i/N_j
              \Bigr\vert_{m_\bot\geq m_\bot^{\rm cut}}$}
              \hfil && \hfil (a) ``thermal"\phantom{$\Bigg\vert$} \hfil
&&                        \hfil (a) ``thermal" \hfil &&
                       \hfil (b) ``flow" \hfil &&
                       \hfil (a) ``flow" \hfil &\cr
          && \hfil &
            & \hfil $m_\bot^{\rm cut}=1.72$ GeV \hfil &
            & \hfil $m_\bot^{\rm cut}=1.2$ GeV \hfil &
            & \hfil $m_\bot^{\rm cut}=1.72$ GeV \hfil &
            & \hfil $m_\bot^{\rm cut}=1.2$ GeV\phantom{$\Bigg\vert$}
         \hfil & \cr
\tablerule
           &&\hfil $\overline{\Xi^-}/\Xi^-$ \hfil &&
             \hfil \phantom{$\Big\vert$} $0.390^*$  \hfil & &
             \hfil --- \hfil & &
             \hfil $0.390^*$  \hfil & & \hfil --- \hfil & \cr
           &&\hfil $\overline{\Lambda}/\Lambda$ \hfil &&
             \hfil \phantom{$\Big\vert$} $0.130^*$  \hfil & &
             \hfil 0.135 \hfil & &
             \hfil $0.130^*$  \hfil & & \hfil 0.133 \hfil & \cr
           &&\hfil $\Xi^-/\Lambda$ \hfil &&
             \hfil \phantom{$\Big\vert$} $0.200^*$  \hfil & &
             \hfil --- \hfil & &
             \hfil $0.200^*$  \hfil & & \hfil --- \hfil & \cr
           &&\hfil $\overline{\Xi^-}/\overline{\Lambda}$ \hfil &&
     \hfil \phantom{$\Big\vert$} $0.600^*$ \hfil & &
             \hfil --- \hfil & &
             \hfil $0.600^*$  \hfil & & \hfil --- \hfil & \cr
           &&\hfil $\Lambda/p$ \hfil &&
             \hfil \phantom{$\Big\vert$} 0.625\phantom{$^*$}  \hfil & &
           \hfil 0.617 \hfil & &
             \hfil 0.888\phantom{$^*$}  \hfil & & \hfil 0.867 \hfil & \cr
          &&\hfil $\overline{\Lambda}/\overline{p}$ \hfil &&
             \hfil \phantom{$\Big\vert$} 2.044\phantom{$^*$}  \hfil & &
           \hfil 2.084 \hfil & &
             \hfil 2.833\phantom{$^*$}  \hfil & & \hfil 2.802 \hfil & \cr
          &&\hfil $\Omega^-/\Xi^-$ \hfil &&
             \hfil \phantom{$\Big\vert$} 0.498\phantom{$^*$}  \hfil & &
           \hfil --- \hfil & &
             \hfil 0.311\phantom{$^*$}  \hfil & & \hfil --- \hfil & \cr
        &&\hfil $\overline{\Omega^-}/\overline{\Xi^-}$ \hfil &&
   \hfil \phantom{$\Big\vert$} 1.502\phantom{$^*$}  \hfil & &
             \hfil --- \hfil & &
             \hfil 0.944\phantom{$^*$}  \hfil & & \hfil --- \hfil & \cr
        &&\hfil $\overline{p}/p$ \hfil &&
             \hfil \phantom{$\Big\vert$} 0.040\phantom{$^*$}  \hfil & &
           \hfil 0.040 \hfil & &
             \hfil 0.041\phantom{$^*$}  \hfil & & \hfil 0.041 \hfil & \cr
          &&\hfil $K_{\rm s}^0/\Lambda$ \hfil &&
             \hfil \phantom{$\Big\vert$} 0.086\phantom{$^*$}  \hfil & &
           \hfil 0.077 \hfil & &
             \hfil 0.138\phantom{$^*$}  \hfil & & \hfil 0.152 \hfil & \cr
          &&\hfil $K_{\rm s}^0/p$ \hfil &&
             \hfil \phantom{$\Big\vert$} 0.054\phantom{$^*$}  \hfil & &
           \hfil 0.048 \hfil & &
             \hfil 0.123\phantom{$^*$}  \hfil & & \hfil 0.132 \hfil & \cr
\tablerule}
}}
\vspace{-0.2cm}
\caption{\tenrm
Input values (${}^*$) and predicted particle ratios for
the WA85 experiment. The parameters for the ``thermal" and ``flow"
scenarios are as in Table 1.}
\vspace{-0.cm}
\end{table}
\pagebreak[5]

$\ $

\vspace{1.2cm}
\begin{figure}[ht]
\vspace{-4cm}
\begin{minipage}[t]{0.475\textwidth}
\centerline{\hspace{0.2cm}\psfig{figure=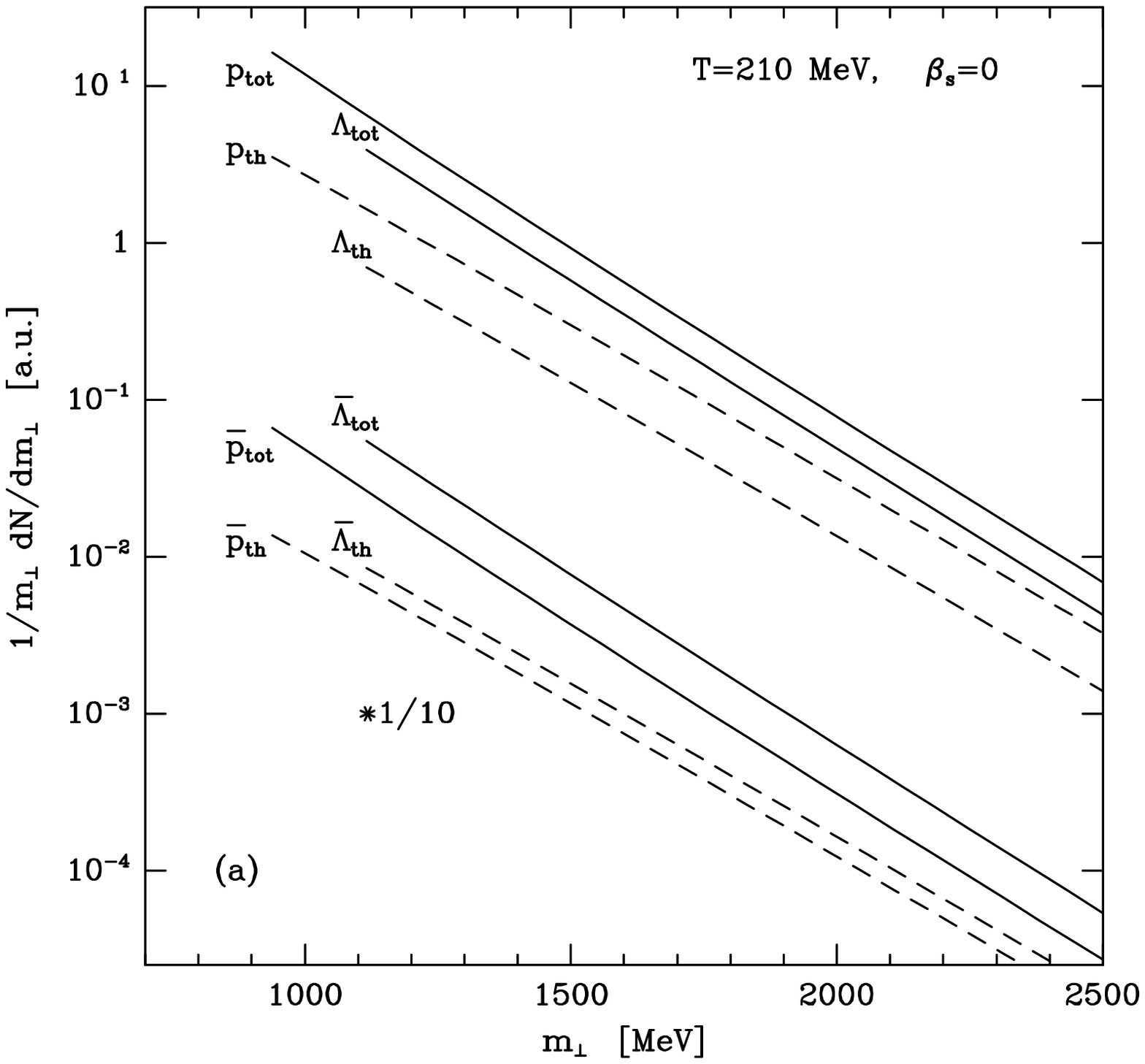,height=11.5cm}}
\end{minipage}\hfill
\begin{minipage}[t]{0.475\textwidth}
\centerline{\hspace{0.2cm}\psfig{figure=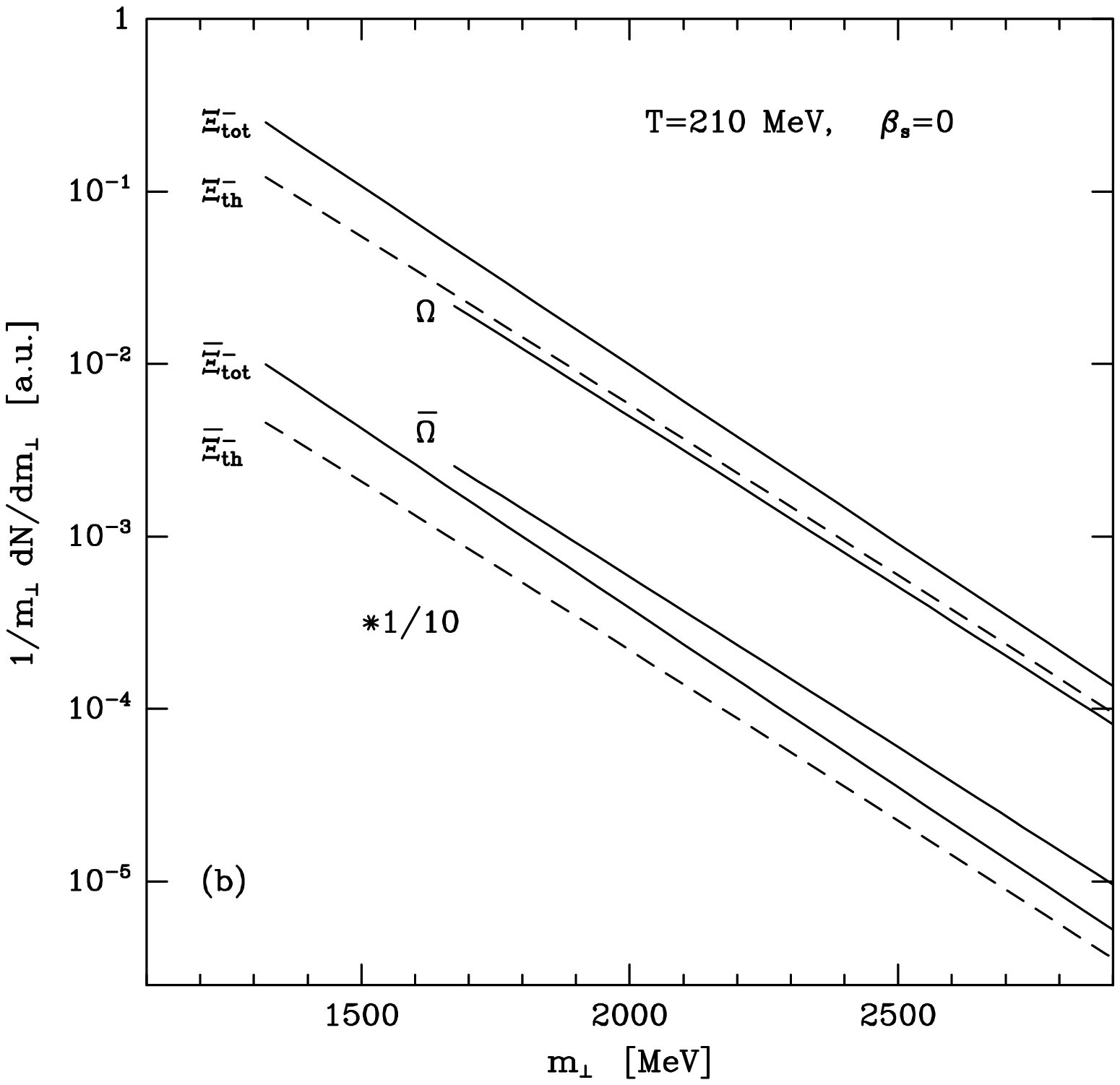,height=11.5cm}}
\end{minipage}
\vspace{ -3.6cm}
\begin{minipage}[t]{0.475\textwidth}
\centerline{\hspace{0.2cm}\psfig{figure=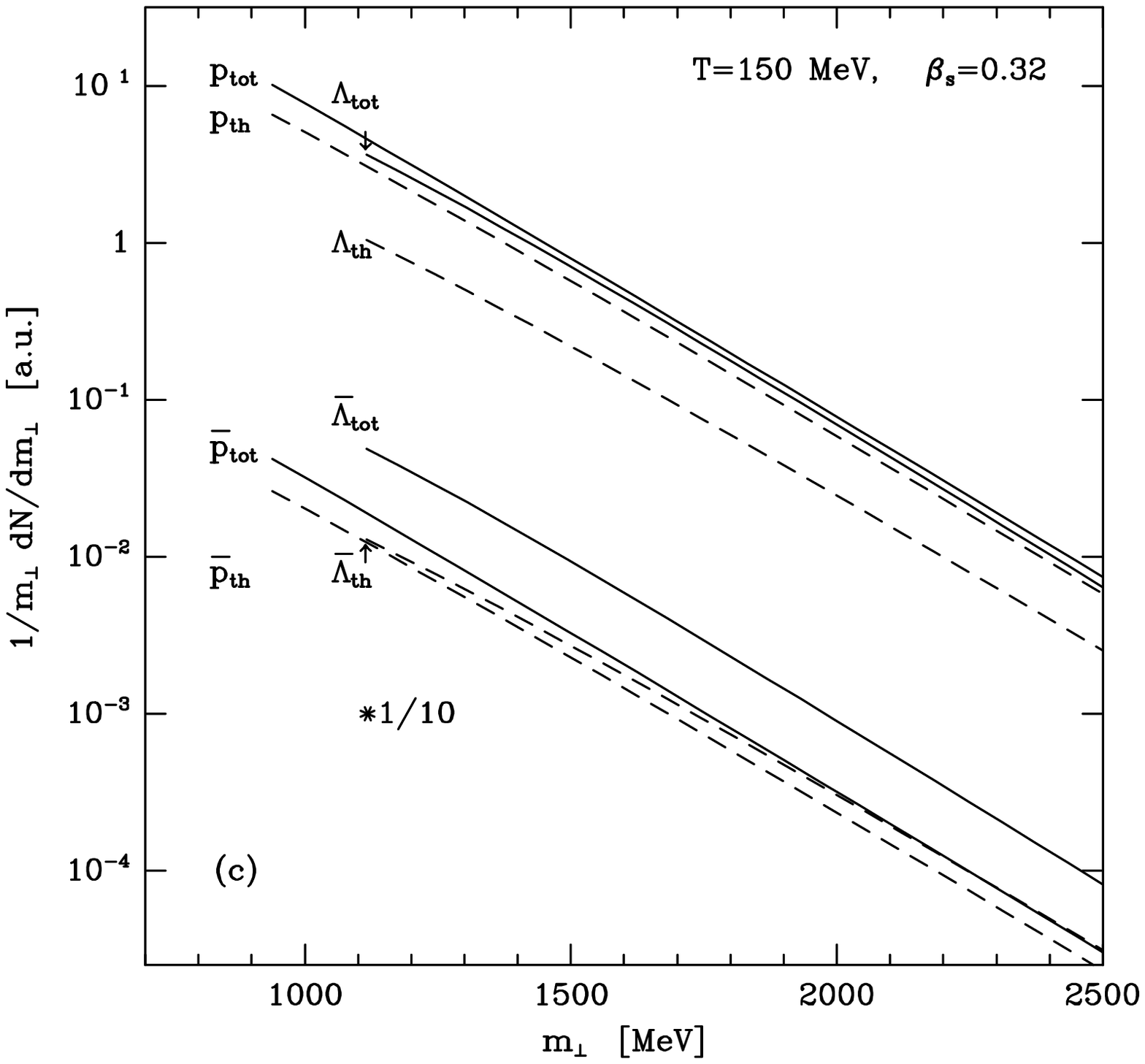,height=11.5cm}}
\end{minipage}\hfill
\begin{minipage}[t]{0.475\textwidth}
\centerline{\hspace{0.2cm}\psfig{figure=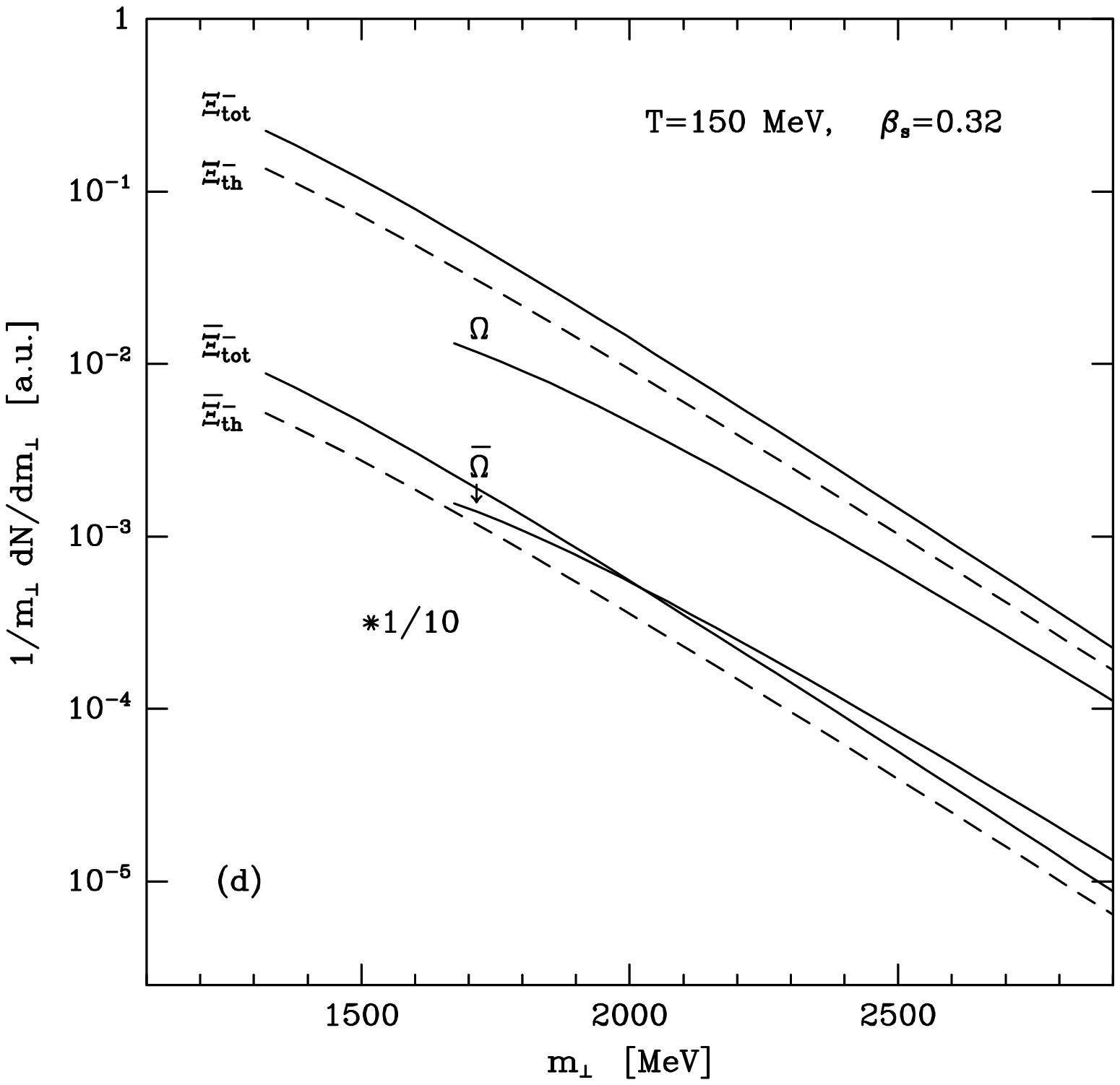,height=11.5cm}}
\end{minipage}
\vspace{ -2.4cm}
\caption{\tenrm
$m_\perp$-spectra of various baryons and anti-baryons
for~the~``thermal" scenario (a,b) and~for~the~``flow" scenario (c,d).
 The overall normalization
of the spectra is arbitrary, their relative normalization is
according to the fugacities from Table 1. Dashed lines indicate direct
thermal production, solid lines give total yields including the
contributions from resonance decays. For
parameters and details see Table 1 and text.
\label{F5new} }
\end{figure}
\vspace{-0.2cm}

temperature, the estimate for $\gamma_{\rm s}$ goes up even higher:
for $T=150$ MeV, $\beta_\bot=0.32$ ({\it i.e.} values which are
consistent with the kinetic freeze-out criterion developed in
\cite{SH92}) we find $\gamma_{\rm s} = 0.9$, {\it c.f.} Table 1,
corresponding to nearly complete saturation of the strange phase
space!
\item[3.] The conclusion that the WA85 data on strange baryon and
anti-baryon production in 200 GeV\,A S--W collisions seem to indicate
a vanishing strange quark chemical potential is extremely stable with
respect to resonance decay contributions and the possibility of a flow
component in the spectra. It will serve as the cornerstone of the
following studies in section\,4.

\end{itemize}
\pagebreak[5]

$\ $

\vspace{1.2cm}

\vspace{-0.2cm}

\subsection{Particle ratios}
\label{3.6}

It is clear that the extraction of three thermodynamic parameters
($\lambda_{\rm q}$, $\lambda_{\rm s}$, and $\gamma_{\rm s}$) from
three independent particle ratios cannot be considered a very
convincing test of the thermal picture, in particular since the
extracted values depend on the choice of the temperature whose
independent derivation from the shape of the $m_\bot$-spectrum is not
quite unique (unless one accepts the theoretical consistency arguments
of Ref. \cite{SH92} which prefer the lower temperature of around 150
MeV). It is therefore mandatory that the picture receives further
experimental tests in the form of other independent particle ratios.

In Table 2 and Figure \ref{F6new} we have therefore included
predictions for several additional particle ratios, in particular
involving $\Omega, \overline{\Omega}$ and $p,\bar p$ which are
expected to become soon available from the WA94 collaboration
\cite{WA94} in a kinematic region compatible with the WA85 results.
Our calculations explore the consequences of both the ``thermal" and
``flow" scenarios above, and anticipate also a possible improvement of
the experimental acceptance towards lower transverse momenta, by
calculating in Table 2 some of the ratios also with an $m_\bot$ cut of
only 1.2 GeV, or plotting them (in Fig.\,\ref{F6new}) as a function of
this cut.

Comparing $m_\bot$-integrated particle ratios with different $m_\bot$
cuts gives an idea of the effects coming from the somewhat different
slopes of thermally emitted particles and resonance decay products.
{}From Fig.\,\ref{F6new} we see that these effects are generally small
for ratios involving only baryons: due to their heavy mass, the slope
difference of parent and daughter particles in resonance decays is
small \cite{sollfrank,SH92}, resulting in a slight steepening of the
spectra at low and intermediate $m_\bot$.  For kaons, on the other
hand, the decay contributions are considerably steeper than the
thermal one, leading to an appreciable upward curvature of the
spectrum at low $m_\bot$, while flattening out to the asymptotic
apparent temperature for $m_\bot >$ 1--1.5 GeV, where the thermal
component dominates. In consequence, as seen in Fig.\,\ref{F6new}, the
$K/\Lambda$ and $K/p$ ratios are proportionally more sensitive to the
$m_\bot$ cut.

It should be kept in mind that the large $\Omega/\Xi^-$ and (in
particular) $\overline{\Omega} / \overline{\Xi^-}$ ratios are partly
due to the additional spin degeneracy of the spin-3/2 $\Omega$ baryon.
The apparent strong $m_\perp^{\rm cut}$-dependence of all ratios
involving $\Omega$ or $\overline{\Omega}$ baryons (especially in the
flow scenario) should be taken with some caution: it results mostly
from the absence of resonance decay contributions to the $\Omega$
spectrum in our calculation and might be modified if (so far unknown)
$\Omega^*$ resonances existed and should be included.

When comparing different ratios involving $\Lambda$'s ({\it e.g.} in
order to estimate the systematic of these ratios with increasing
strangeness), one should always keep in mind the considerable
$\Sigma^0$ contribution which, at $T=210$ MeV, effectively changes the
degeneracy factor associated with the $\Lambda$ by about a factor 1.6.
Once this is corrected for, {\it i.e.} the $\Sigma^0$ contribution is
accounted for, one finds the inequalities
 \begin{equation}
    \gamma_{\rm s}^2 >
    {\Lambda\overline{\Lambda} \over p\bar p} >
    {\Xi^-\overline{\Xi^-} \over \Lambda\overline{\Lambda}} >
    {1\over 4} \,
    {\Omega\overline{\Omega} \over \Xi^-\overline{\Xi^-}}\, .
 \label{gam4}
 \end{equation}
Paradoxically, Eq.\,(\ref{gam4}) suggests an increasing importance of
resonance decay corrections with increasing strangeness content, as
compared to the estimate, Eq.\,(\ref{gam3}). This is due to the fact
that the stranger baryons are also heavier and thus receive less
resonance contributions (because the corresponding excited states are
strongly suppressed by the mass term in the Boltzmann factor) than the
lighter ones (whose resonances are easier to excite). In part this
tendency may be somewhat exaggerated by our cut in the resonance
spectrum at about 2 GeV; taking into account even higher resonances
(although it is not clear whether can indeed be excited with
statistical probabilities in nuclear collisions) might modify in
particular the $\Omega/\Xi^- $ ratios.

We regard the ratios\footnote{Ratios involving protons should be taken
with a grain of salt since it may not be easy experimentally to remove
the contamination by cold projectile and target spectator protons.
Similarly, those ratios which involve kaons may come out differently
than predicted here because, as we argue in section \ref{sect5}, the
assumption of chemical equilibrium may break down for kaons if they
are created by hadronization of a QGP.} predicted in Table 2 and
Fig.\,\ref{F6new} as a qualitative test of the
idea that the emitter is a fireball which is in thermodynamic
equilibrium with the given thermodynamic parameters. However, even if
this test is eventually passed, the fact that we can successfully
describe all observed particle ratios in this way does not yet tell us
immediately the nature of the emitting source (HG or QGP). That
information resides in supplementary information concerning the global
properties of the source and their {\it relationship} to the extracted
thermodynamic parameters, as well as in the systematic behavior of
these parameters under varying experimental conditions (collision
energy, projectile and target size, collision centrality etc.). These
important issues will be discussed in more detail in the following
sections \ref{sect4} and \ref{sect5}.

\subsubsection{Relation between strange baryon ratios from the
constrained fireball}

Since thermal particle ratios are functions of only $\mu_{\rm B}$ and
$\mu_{\rm s}$, the accidental vanishing of $\mu_{\rm s}$ at a
temperature $T\simeq$ 200--220 MeV due to the constraint to
strangeness neutrality removes the opportunity to simply distinguish
the two phases --- a HG phase of strongly interacting matter, at this
level of discussion, is {\em indistinguishable} from QGP as long as a
measurement of the relative multiplicities of strange particles is
considered at one, and only one collision energy. Even in this case
many theoretical questions arise, such as why is $\mu_{\rm s}=0$ or
why is $\gamma_{\rm s}$ near to unity? However, as Figs.\,\ref{F2}a,
\ref{F2}b, \ref{F2}c clearly show, this will not remain a problem at
higher or lower temperature. To make this point more quantitative, we
now study strange baryon ratios arising at different temperatures from
thermal fireballs {\em constrained} to a (nearly) vanishing
strangeness.

\begin{figure}[t]\vspace{-2cm}
\begin{minipage}[t]{0.475\textwidth}
\centerline{\hspace{0.2cm}\psfig{figure=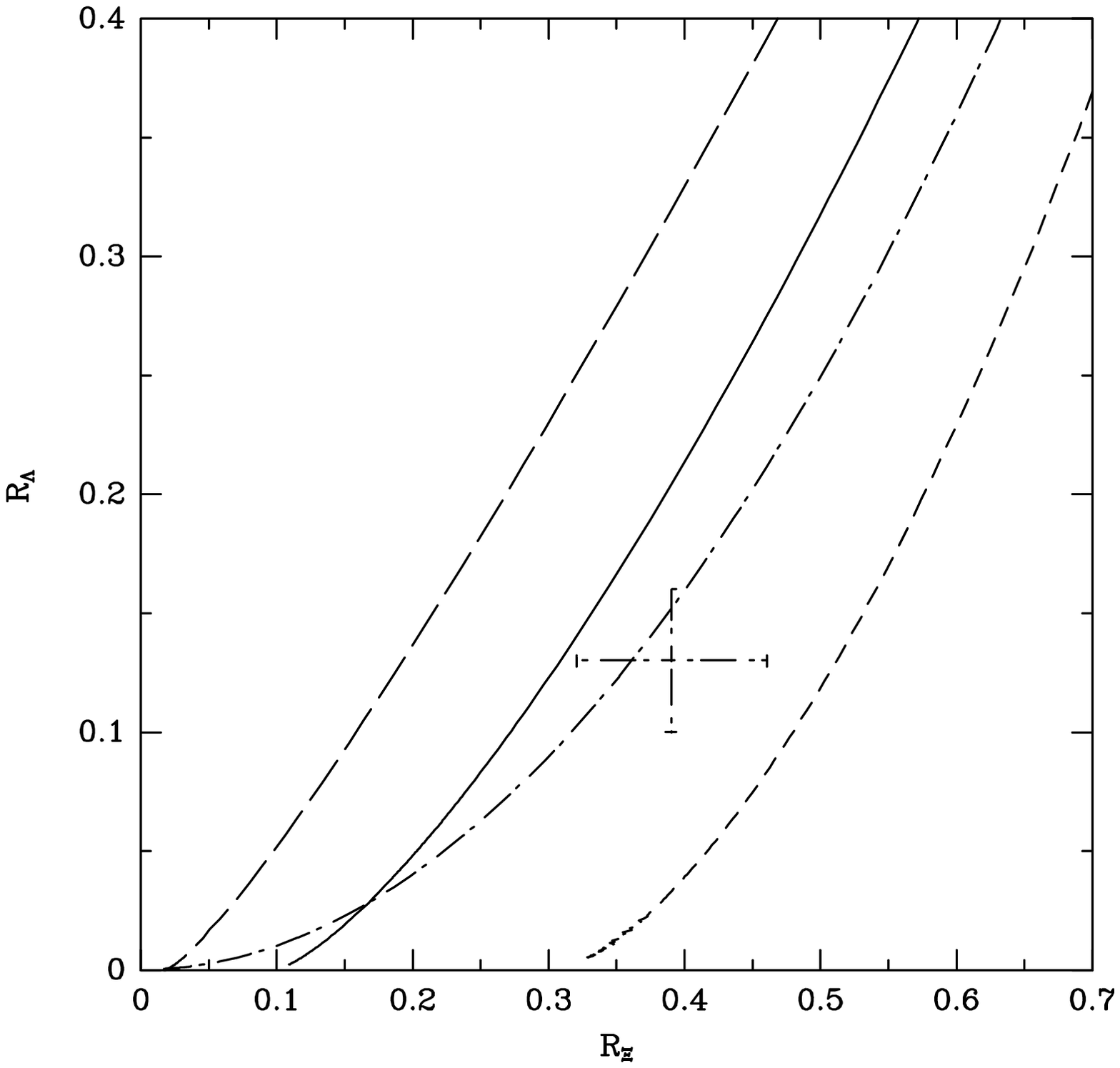,height=11.5cm}}\vspace{-2cm}
\caption{\tenrm
$R_\Lambda$ versus $R_\Xi$. The long-dashed line corresponds to $T = 150$
MeV, the solid line to $T = 200$ MeV, and the dashed line to $T = 300$
MeV in the HG. The dashed-dotted line corresponds to QGP ($\mu_{\rm s}\equiv
0$).}\protect\label{F3}
\end{minipage}\hfill
\begin{minipage}[t]{0.475\textwidth}
\centerline{\hspace{0.2cm}\psfig{figure=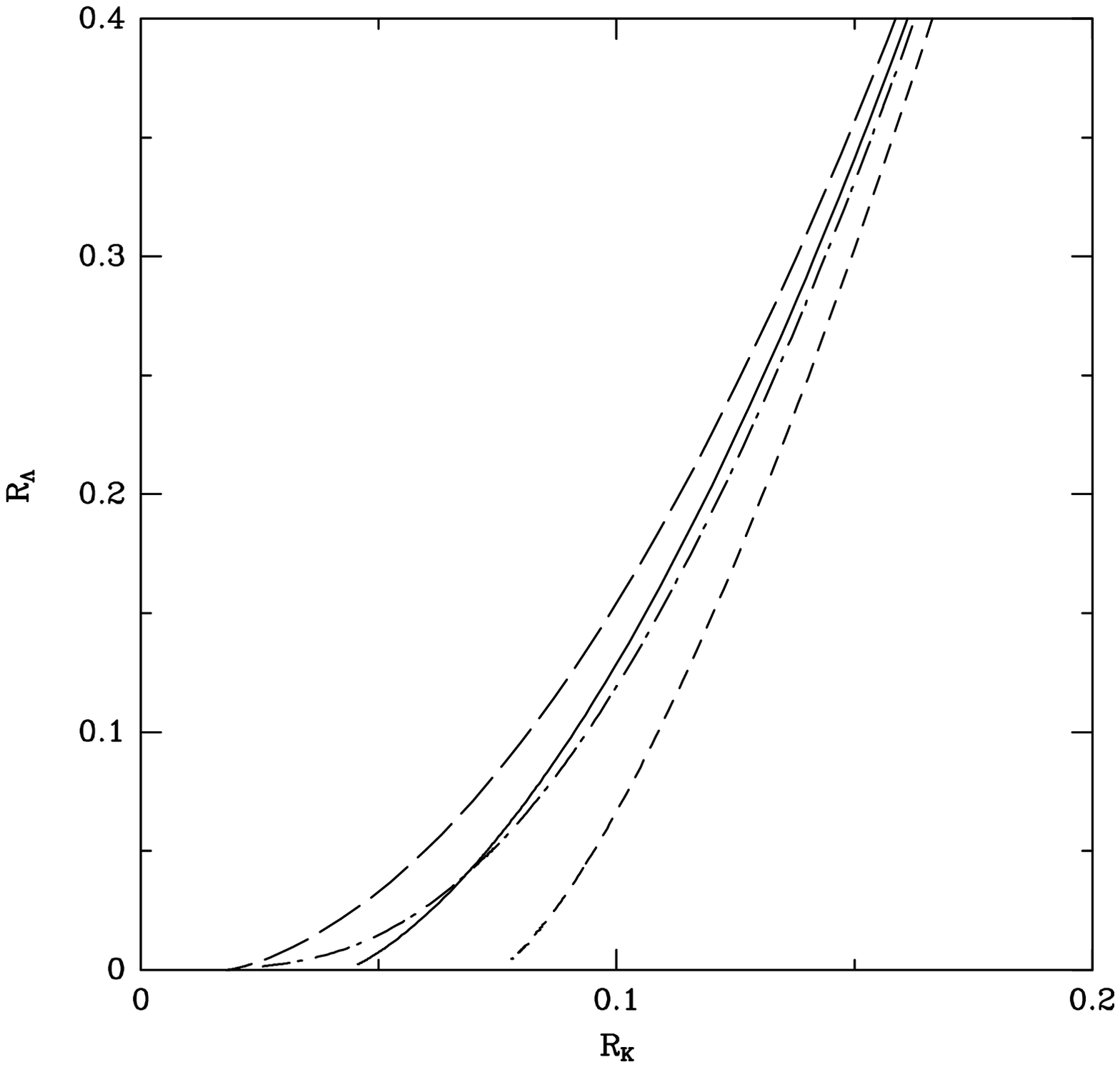,height=11.5cm}}\vspace{-2cm}
\caption{\tenrm  $R_\Lambda$ versus $R_{\rm K}$ with the same conventions as in
Fig.\,\protect\ref{F3}.
}\protect\label{F4}
\end{minipage}
\end{figure}

In Fig.\,\ref{F3} we show, for the case of exactly vanishing
strangeness and for various temperatures, the resulting relation
between $R_\Xi$ and with $R_\Lambda$. In addition to the HG results
for temperatures $T=200$ MeV (solid line), $T=150$ MeV (dashed line)
and $T=300$ MeV (dotted line) we show the case $\mu_{\rm s}=0$
corresponding to a QGP source hadronizing rapidly by quark
recombination (dashed-dotted line).  The cross corresponds to the
result reported by the WA85 experiment \cite{WA85}. As can be seen,
the $\mu_{\rm s}=0$ curve nearly coincides with the $T=210$ MeV curve,
as noted before \cite{LTR92,Cley92}, but fireballs decoupling at
significantly lower or higher temperatures (as expected at lower and
higher collision energies, respectively) would lead to significantly
different particle ratios if the thermal model is correct.

\subsubsection{High-$m_\perp$ kaon abundances}
In addition to the baryon ratios we can also consider the ratio of
kaons to hyperons, again at fixed $m_\bot$. Because of the
experimental procedures used, which rely on the observation of the
disintegration of neutral strange particles into two charged decay
products, a comparison of the $K^0_{\rm s}$ with the $\Lambda$ (which
includes the $\Lambda$'s from the electromagnetic decay of
$\Sigma^0$'s) is the most useful next step. We therefore introduce:
 \begin{eqnarray}
    R_{\rm K} \equiv
    {K^0_{\rm s}\over \Lambda+\Sigma^0} =
    {1\over 8}{\lambda_{\rm s}/\lambda_{\rm d} +
               \lambda_{\rm d}/\lambda_{\rm s} \over
               \lambda_{\rm s} \lambda_{\rm u} \lambda_{\rm d}} \, .
 \label{rk}
 \end{eqnarray}
where the second identity is again, of course, only valid if resonance
decay contributions can be neglected. For the rather light kaons there
exist, however, many different possibilities for secondary production
through resonance decays, which limits somewhat the practical
usefulness of this ratio. For example, if we use the thermodynamic
parameters extracted in subsection~\ref{3.4} from the WA85 data, we obtain
for the ``thermal" interpretation of the $m_\bot$-slope ($T=210$ MeV,
no flow, $m_\bot=1.72$ GeV) a value $R_K=0.086$, if all resonance
contributions are included, while the ground state contributions alone
according to Eq.~(\ref{rk}) would give $R_K = 0.102$ ($R_K=0.125$ if
the $\Lambda$--$\Sigma^0$ mass difference is taken into account).
Since the size of these corrections depends on the fireball
temperature and the $m_\bot$-cut, a full quantitative study does
presently not appear practical.

Given these uncertainties, and in order to demonstrate the
insensitivity of the $K/\Lambda$-ratio to the nature of the fireball,
we show in Fig.\,\ref{F4} (as a matter of principle) the
uncorrected ratio $R_K$ as a function of the uncorrected ratio
$R_\Lambda$: as the figure clearly shows, this observable is rather
insensitive to both the temperature of the source and to its intrinsic
structure ($\mu_{\rm s}=0$ or $\mu_{\rm s}\ne0$). The lines are as in
Fig.\,\ref{F3}, and we note again the near coincidence of the $T=200$
MeV HG result with the QGP result. But even as $T$ is changed to 150
or 300 MeV, the variation of $R_\Lambda$ at fixed $R_{\rm K}$ stays
well within the present experimental error bars and will thus not
permit to identify the properties of the source.

\section{Total particle multiplicity and entropy}
\label{sect4}

We now return to discuss in detail a very important observation which
is related to the particle multiplicity and charge flow associated
with strange baryon production \cite{Let92,LTR92}. Even though no data
are yet available on multiplicity densities in coincidence with
strange particle production, the properties of the HG and QGP
fireballs with regard to their entropy content are so drastically
different that we can draw at least qualitative conclusions by
comparing multiplicity data from S--Pb collisions with S--W
strangeness production, despite the fact that the two targets
($^{207}$Pb and $^{184}$W) differ slightly in mass.

QGP and HG states are easily distinguishable in the regime of values
$\mu_{\rm B},\ T,\ \gamma_{\rm s}$ of interest here. The entropy per
baryon in the HG is ${\cal S}^{\rm HG}/{\cal B} = 21.5 \pm 1.5$.
Consequently, the pion multiplicity which can be expected from such a
HG fireball is at best $dN/dy=4 \pm 0.5$. This is less than half of
the QGP based expectation, which we can easily estimate by considering
the perturbative QGP equation of state: Up to corrections of order
$(\mu_{\rm B}/\pi T)^2 \sim 0.03$, the leading term of the specific
entropy from light quarks and gluons is:
 \begin{eqnarray}
   {{\cal S}^{\rm QGP}\over{\cal B}}&&\hspace{-0.6cm}=
   {{3\pi^2}\over2}\Big({T\over\mu_q}\Big)\Big[
   {{14(1-50\alpha_s/21\pi)}\over{15(1-2\alpha_s/\pi)}}+
   {{32(1-15\alpha_s/4\pi)}\over{45(1-2\alpha_s/\pi)}}
   \Big]
 \nonumber \\
  &&\hspace{-0.6cm}= 17\, (T/\mu_q)\quad \mbox{ for }\alpha_s=0.6\quad .
 \end{eqnarray}
Adding to this the considerable entropy content of the strange quarks
(neglecting for technical reasons the $\alpha_s$-corrections in this
case), we find for the QGP a specific entropy of about 45 units at the
same value of $T/\mu_{\rm q}$. Poor knowledge of the value of
$\alpha_s$ and the strange quark mass in the QGP phase make this
estimate uncertain by about $\pm15$\%. Still, the difference to the HG
result is considerable in terms of experimental sensitivity. Checking
the theoretical sensitivity we note that the point at which the
entropy of HG and QGP coincide {\it and} strangeness vanishes {\it
and} $\lambda_{\rm s}\simeq1$ is at $T \simeq 135$ MeV, $\mu_{\rm B}
\simeq 950$ MeV, quite different from the region of interest here. We
therefore consider now an observable which is capable to distinguish
between the HG and QGP via their entropy content.

\subsection{EMU05 data}

\begin{figure}[t]\vspace{-2.2cm}
\begin{minipage}[t]{0.475\textwidth}
\centerline{\hspace{0.2cm}\psfig{figure=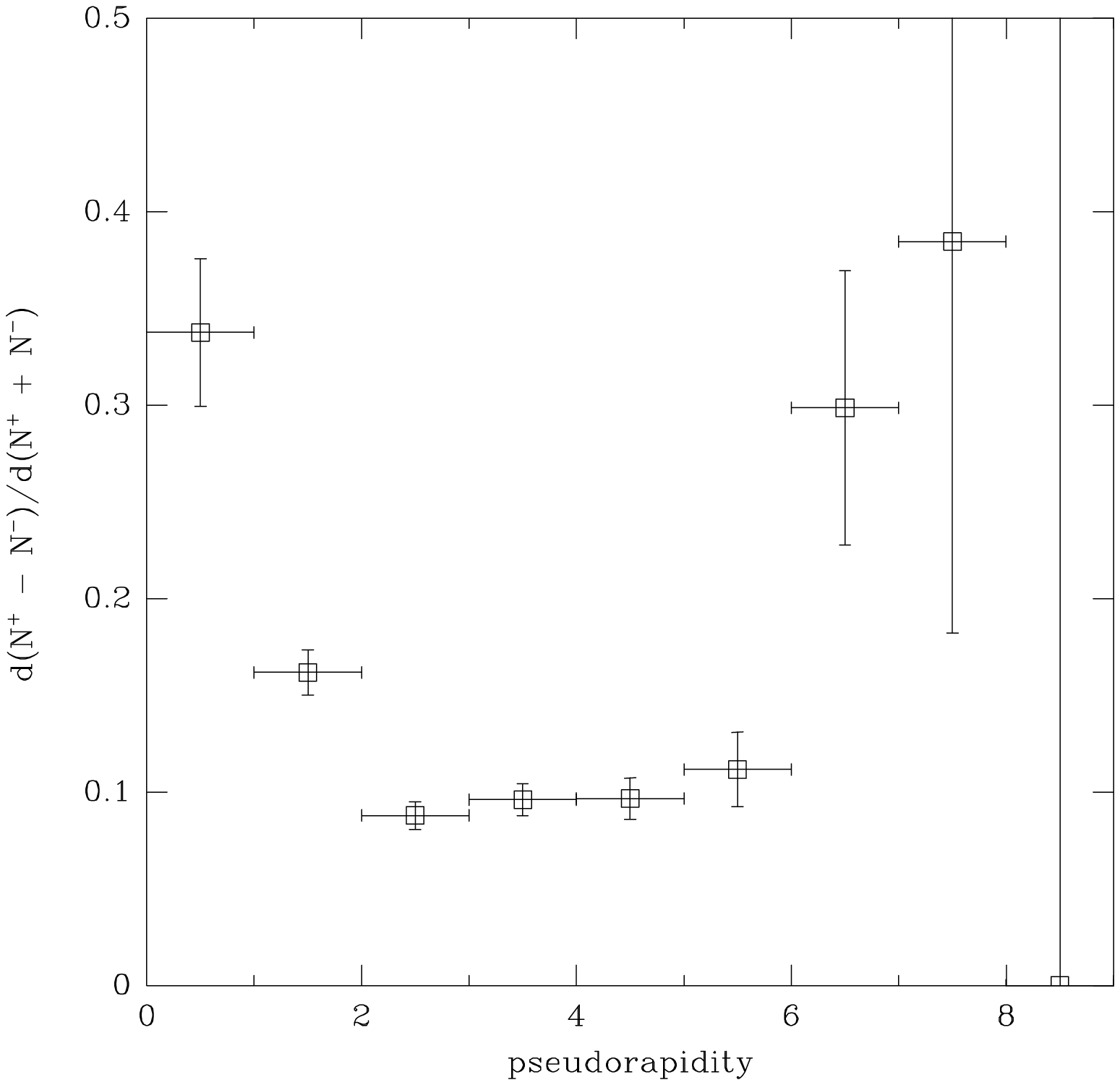,height=11.5cm}}\vspace{-2cm}
\caption{\tenrm
Emulsion data for charged particle multiplicity as function of
pseudo-rapidity: the difference of positively and negatively charged
particles normalized by the sum of both polarities. (Curtesy of
EMU05 collaboration, Y. Takahashi et al. \protect\cite{EMU05}).}
\protect\label{F5}
\end{minipage}\hfill
\begin{minipage}[t]{0.475\textwidth}
\centerline{\hspace{0.2cm}\psfig{figure=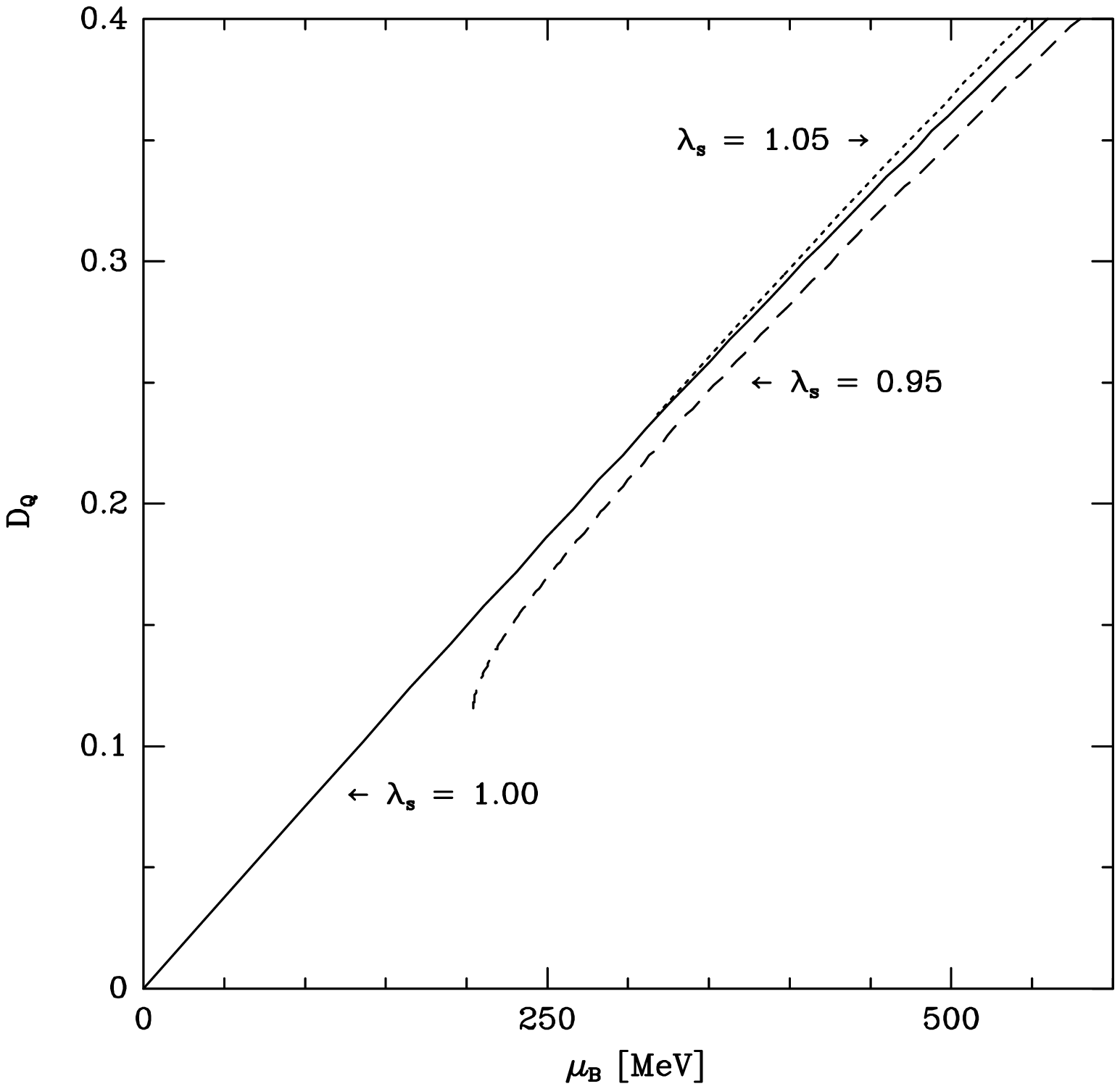,height=11.5cm}}\vspace{-2cm}
\caption{\tenrm  $D_{\rm Q}$ as a function of $\mu_{\rm B}$ for fixed
$\lambda_{\rm s}=1\pm0.05$ and conserved zero strangeness in HG.
}\protect\label{F6}
\end{minipage}\\
\begin{minipage}[c]{0.475\textwidth}
\vspace{2.5cm}
\caption{\tenrm
Entropy per baryon ${\cal S/B}$ as a function of $D_{\rm Q}^{-1}$ for
fixed $\lambda_{\rm s}=1$ and conserved zero strangeness. The
lower, dashed curve does not take into account the strange hadrons.}
\protect\label{F7}
\end{minipage}\hfill
\begin{minipage}[c]{0.475\textwidth}
\vspace{-3.1cm}
\centerline{\hspace{0.2cm}\psfig{figure=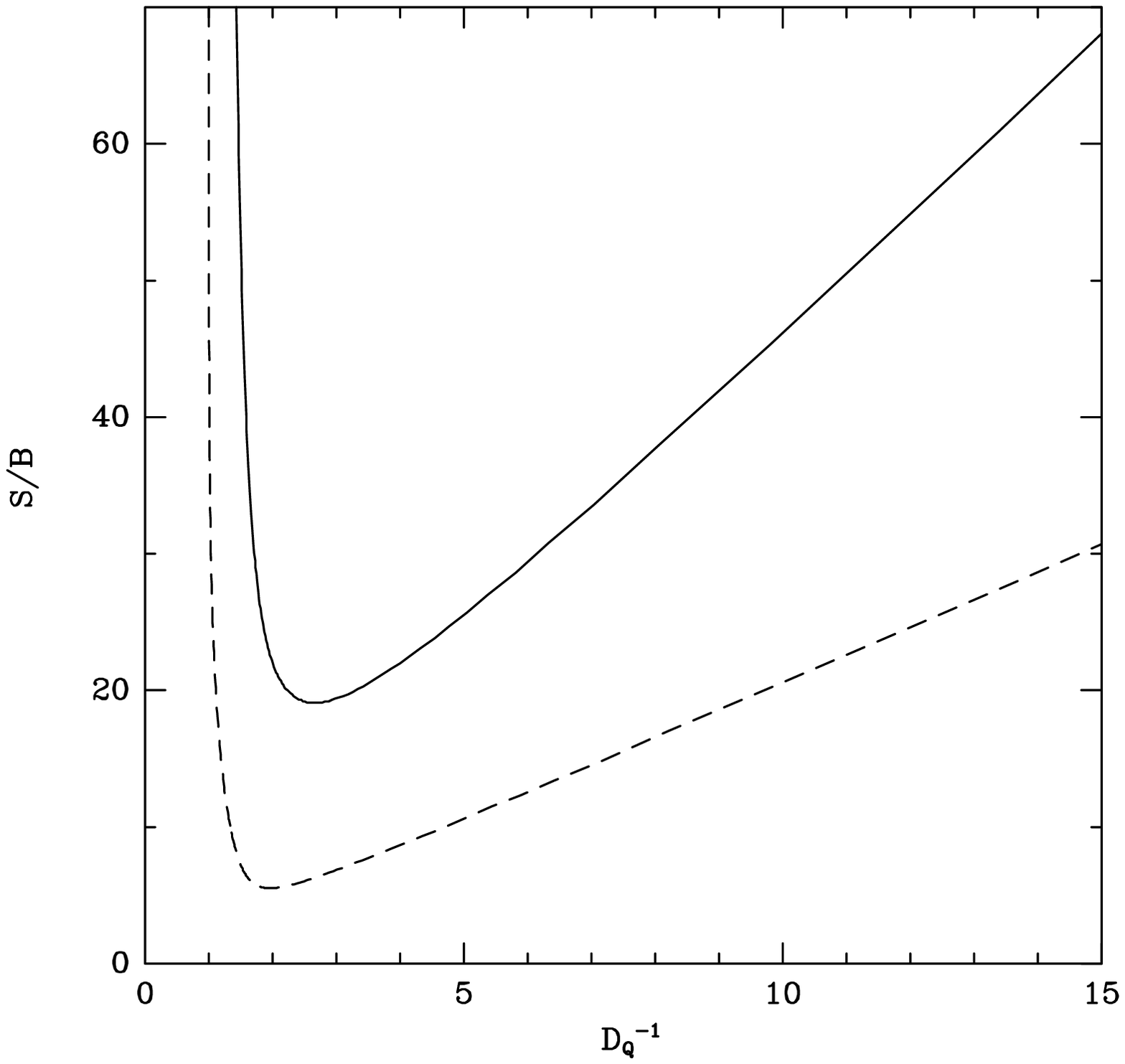,height=11.5cm}}
\vspace{-2.7cm}

\end{minipage}\end{figure}

Total charged particle multiplicities (excluding target/projectile
fragments) {\it above} 600 in the  central region have been observed
by EMU05 \cite{EMU05} in 200 GeV\,A S--Pb collisions, corresponding
possibly to a total particle multiplicity of as large as 1000. We
believe that such large multiplicities are implied by a QGP scenario
for the central fireball, while being hardly compatible with the
corresponding HG interpretation. The data of the EMU05 collaboration
\cite{EMU05} were obtained with a high multiplicity ``trigger'' which
makes them useful in comparison with the WA85 results which are also
triggered on high multiplicity. Some of these charged multiplicity
data from emulsions are shown as a function of rapidity in
Fig.\,\ref{F5}. We depict the quantity $D_{Q}$, the difference in the
number of positively and negatively charged particles normalized by
their sum \cite{EMU05}:
 \begin{equation}
   D_{\rm Q} \equiv {{N^+ - N^-} \over {N^+ + N^-}}
  \label{DQ1}
 \end{equation}

The data points correspond to 15 fully scanned ``central" events of
200 GeV\,A S--Pb interactions, with the trigger requirement being a
total charged multiplicity $> 300$. Reaction spectators (target
fragments) are not observed in this experiment. At central rapidity a
value of 0.08--0.09 is found for $D_{\rm Q}$. We note the rise of
$D_{\rm Q}$ in the projectile and target pseudorapidity regions,
indicating a more rapid decrease of the particle multiplicity (mostly
pions) than of the positive particle excess (protons). Obviously this
feature proves the partial transparency at CERN energies of the
Pb nucleus to the incoming sulphur projectile; it will be quite
interesting to see if this feature persists for heavier projectiles
and in particular how the shape is in Pb--Pb collisions.

\pagebreak[5]

$\ $

\vspace{2.5cm}

$\ $

\vspace{-3.cm}


In our theoretical work we address primarily the central region, in which
the smallest values of $D_{\rm Q}$ indicate the highest multiplicity
production.

\subsection{Significance of $D_{\rm Q}$}
The numerical evaluation of $D_{\rm Q}$ in a HG at $\lambda_{\rm s} =
1 \pm 0.05$, where we fix the temperature for each $\mu_{\rm B}$ such that
strangeness is conserved, is shown in Fig.\,\ref{F6}. We see that
approximately
 \begin{equation}
  D_{\rm Q} \approx {\mu_{\rm B}\over\mbox{1.3 GeV}}\
  \quad \mbox{ for } \quad \mu_{\rm B}<0.6\mbox{ GeV}.
 \label{DQ}
 \end{equation}
This HG result is extremely simple, considering the complexity of the
calculation. We thus see that in the only permissible HG scenario for
the strangeness source which (according to Eq.~(\ref{chempot})) {\em
has to have} $\mu_{\rm B}\sim 0.34$ GeV, Eq.\,(\ref{DQ}) would predict
a value of $D_{\rm Q}\sim 0.26$; this is incompatible with the EMU05
results, see Fig.\,\ref{F5}, where the central value is $D_{\rm
Q}(y=2.5 \pm 0.5) = 0.088 \pm 0.007$. At this value of $D_{\rm Q}$ the
HG phase would have $\mu_{\rm B}=115$ MeV, rather than the $340$ MeV
we have extracted from the strange baryon data. Despite the fact that
both measurements use different targets we believe that we can draw
the conclusion that a HG fireball is incompatible with the combined
EMU05 and WA85 data.

\begin{figure}[ht]\vspace{-5cm}
\begin{minipage}[t]{0.475\textwidth}
\centerline{\hspace{0.2cm}\psfig{figure=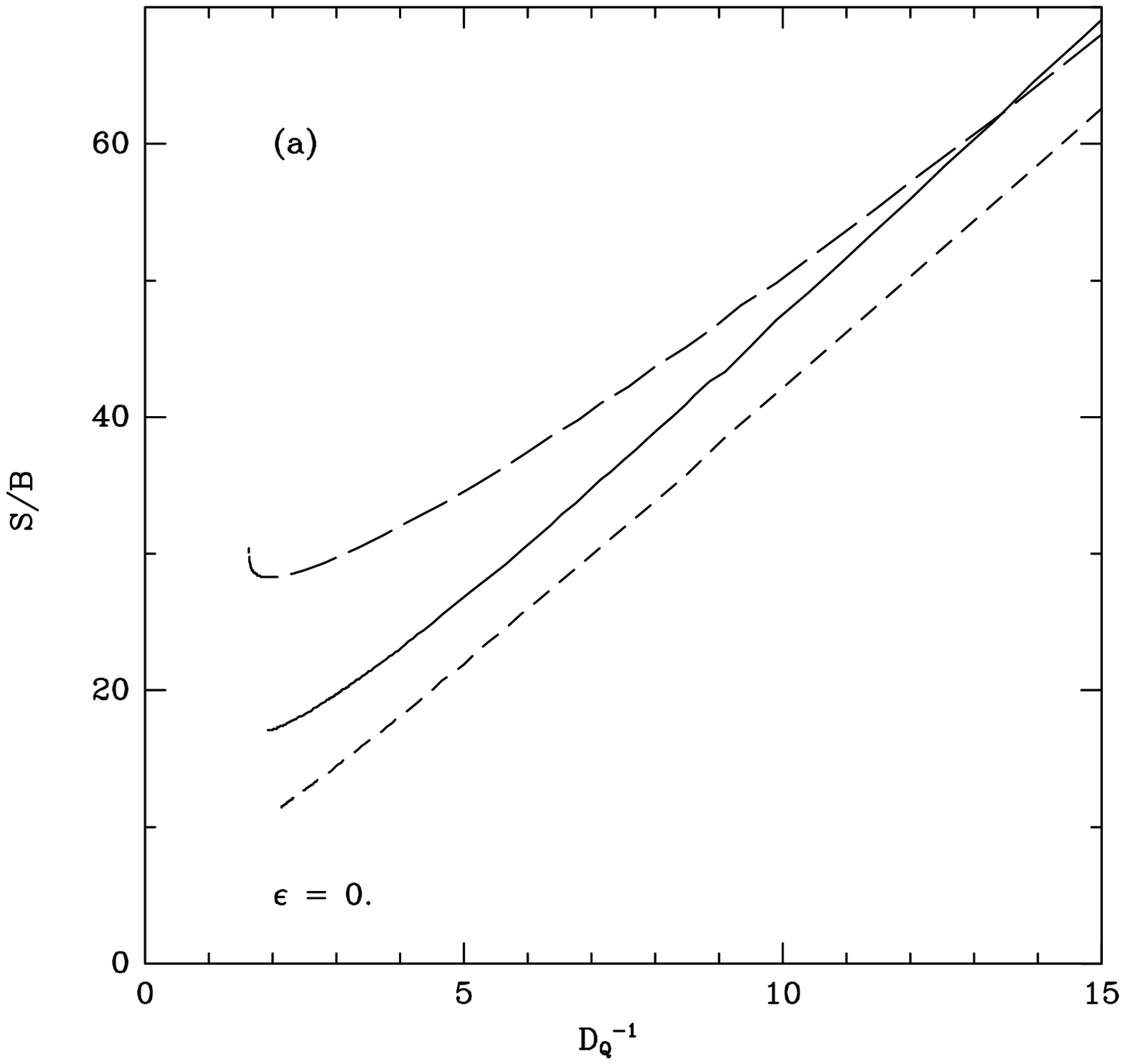,height=11.5cm}}
\vspace{-2.2cm}
\end{minipage}\hfill
\begin{minipage}[t]{0.475\textwidth}
\centerline{\hspace{0.2cm}\psfig{figure=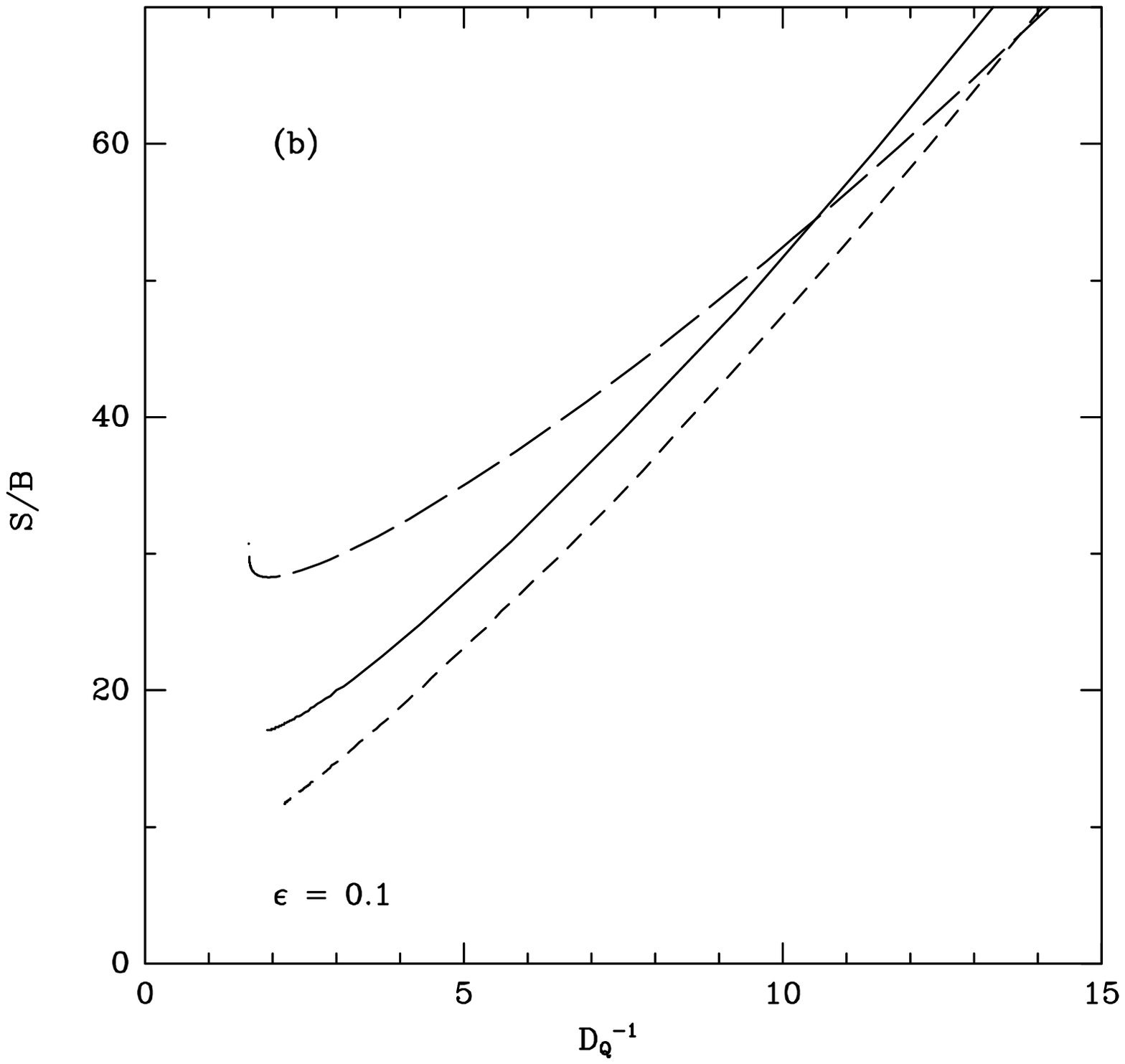,height=11.5cm}}
\vspace{-2.2cm}
\end{minipage}\\
$\ $

\vspace{-1.4cm}
\begin{minipage}[c]{0.475\textwidth}
\caption{\tenrm
a) Entropy per baryon as a function of $D_{\rm Q}^{-1}$
at fixed temperature
($T=$ 150, 200, 300 MeV) for conserved zero strangeness (same conventions
as in Fig.\,\protect\ref{F2}a).
b) and  c) Same as Fig.\,\protect\ref{F8}a but for
$\epsilon\pm 0.1$ (compare to Fig.\,\protect\ref{F2}b, c).\protect\label{F8}
}
\end{minipage}\hfill
\begin{minipage}[c]{0.475\textwidth}
\vspace{-2.2cm}
\centerline{\hspace{0.2cm}\psfig{figure=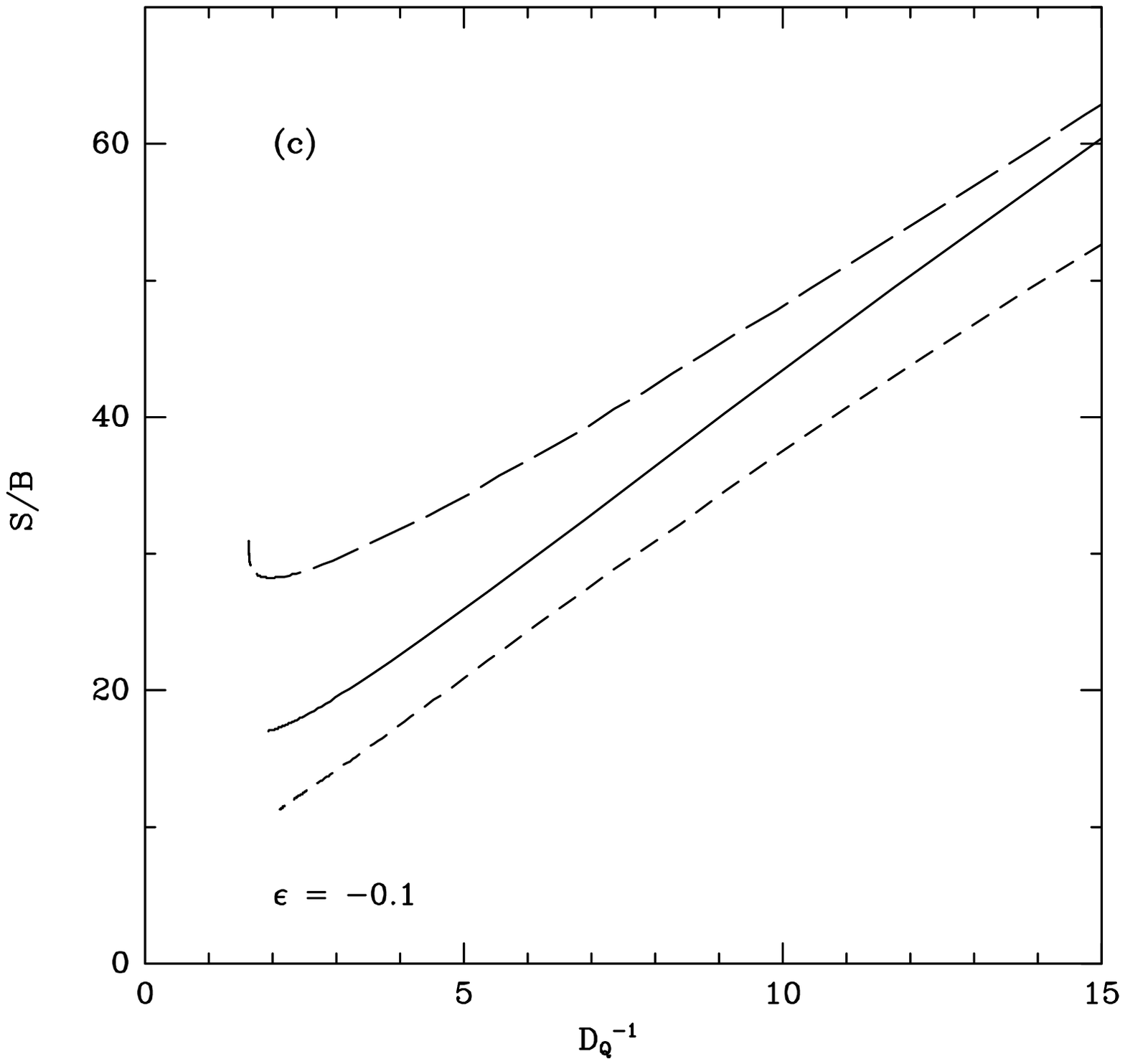,height=11.5cm}}
\vspace{-2.2cm}
\end{minipage}

\end{figure}

Physically the variable $D_{\rm Q}$ represents essentially the inverse
of the specific entropy. The measured very small value of $D_{\rm Q}$
thus implies a very large value for ${\cal S/B}$. In Fig.\,\ref{F7} we
show the specific entropy as a function of $D_{\rm Q}^{-1}$ for a
zero-strangeness hadron gas at $\lambda_{\rm s} = 1$. We see the very
strong direct correlation between these two quantities. We supplement
this observation in Figs.\,\ref{F8}a, \ref{F8}b, and \ref{F8}c by also
showing ${\cal S/B}$ as a function of $D_{\rm Q}^{-1}$ at fixed $T=$
150, 200, 300 MeV, varying $\epsilon = 0,\,\pm0.1$. We observe that,
within the range of these parameters, the small observed value of
$D_{\rm Q}$ always requires a rather large specific entropy ${\cal
S/B} = 50\pm4$. Thus $D_{\rm Q}^{-1}$ effectively measures the entropy
per baryon, independent of the specific thermal conditions of the
source.

\begin{figure}[t]
\begin{minipage}[t]{0.475\textwidth}
$\ $

\subsection{NA35 data}

Even though the use of thermal model is less convincing for the small
collision system S--S, we will shortly consider the results obtained by
the NA35 collaboration \cite{NA35,Wenig} in S--S interactions at 200
GeV\,A. In this case we can base our discussion on the reported charged
particle rapidity densities \cite{Wenig} and the measured
$\overline{\Lambda} / \Lambda$ ratio \cite{NA35}. S--S interactions show
at central rapidity visibly less baryon number stopping than S--W
collisions; from the data on $d(N^+-N^-)/dy$ and $dN^-/dy$ in
\cite{Wenig} we read off a central rapidity value of $D_{\rm Q}^{\rm
S+S}(y=3)=0.065$. The raw $\overline{\Lambda} / \Lambda$ ratio
(uncorrected for $\Xi$-decays) is strongly peaked at central rapidity and
rises there to a value $\overline{\Lambda} / \Lambda = 0.34$ (with error
bars of order 50\%) \cite{NA35}. The observed transverse mass spectra can
again be interpreted in terms of a thermal, longitudinally expanding
central

\end{minipage}\hfill
\begin{minipage}[t]{0.475\textwidth}\vspace{-1.6cm}
\centerline{\hspace{0.2cm}\psfig{figure=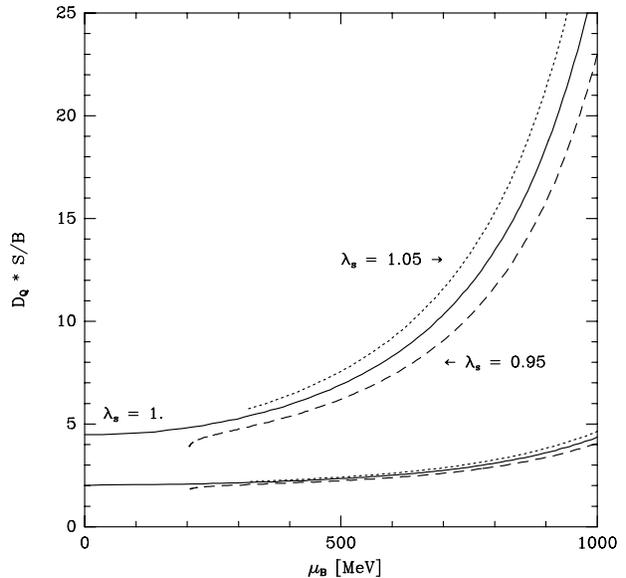,height=11.5cm}}\vspace{-2cm}
\caption{\tenrm  The product $D_{\rm Q}\cdot({\cal S/B})$ as function
of $\mu_{\rm B}$ for
fixed $\lambda_{\rm s}=1\pm0.05$ and zero strangeness (compare to
Fig.\,\protect\ref{F6}). The lower set of curves was obtained without including
strange particles.
}\protect\label{F9}
\end{minipage}
\end{figure}

Please note that in this connection it is essential to correctly
account for the strange particles. To illustrate this point and to
understand the value of $D_{\rm Q}$ we show in Fig.\,\ref{F9} the
product $D_{\rm Q}{\cdot}({\cal S/B})$ as a function of $\mu_{\rm B}$
for fixed $\lambda_{\rm s}=1\pm0.05$, where akin to Fig.\,\ref{F7},
the lower set of curves was obtained without strange particles. We
see that strange particles have considerable impact on the result.

Both the limiting value and near constancy of the result obtained
without strange particles can be easily understood since in this case
$D_{\rm Q}{\cdot}({\cal S/B})$ is effectively the entropy per pion. To
see this note that in absence of strange particles we have (neglecting
the small $u$--$d$ asymmetry and assuming pion symmetry $N_{\pi^+} =
N_{\pi^-} = N_{\pi^0} = N_\pi/3$):
 \begin{equation}
   D_{\rm Q}
   \to 0.75 {{\cal B} \over N_\pi}\ {1 \over
   {1+1.5 \sum_i N_i/N_\pi}}
  \label{DQBP}
 \end{equation}
where the last term in the denominator involves the sum over all initial
charged particles other than pions, in particular heavy mesons, protons
and anti-protons, etc. Considering the product of $D_{\rm Q}$ with $\cal
S/B$, the baryon content cancels and the result is effectively entropy
per pion renormalized (as it turns out numerically with a factor 2) by
the degeneracy factors and resonance contributions. This also explains
why the product is rather constant as $\mu_{\rm B}$ (and along with it
$T$) changes: there is a slow increase in the entropy per pion with
increasing $\mu_{\rm B}$ since the resulting decrease in $T$ enhances
the relative importance of the heavier baryon resonances with the
associated larger entropy per particle.

The question arises whether the QGP fireball model would be in better
agreement with the combined WA85 and EMU05 data, or if the
disagreement of the HG picture with the data is rather an indication
for the failure of the fireball model in general. In order to be able
to answer this question in detail we need in principle to have a
sophisticated model for the hadronization of a QGP. However, as long
the final state is one with thermal momentum distributions, the
relationship between $D_{\rm Q}$ and ${\cal S/B}$ is likely to be
similar to that of Fig.\,\ref{F7} even if the observed hadrons result
from a hadronizing QGP. The reason is that the product $D_{\rm Q}
\cdot ({\cal S/B})$ (in other words, the entropy per pion) will not
deviate drastically from its equilibrium hadron gas value even in a
state which is considerably out of chemical equilibrium. Noting that
there is little entropy production in the QGP
hadronization~process~\cite{FRI83}~we~can

\pagebreak[5]

$\ $

\vspace{2.cm}

\noindent thus subsume that after hadronization the result
would be a point close to the curve shown in Fig.\,\ref{F7}. We see
that the value $D_{\rm Q} \simeq 0.09$ implies ${\cal S/B} \sim$ 50,
which is indeed of the order required by a hadronizing QGP with the
thermal parameters extracted in section 3. We note that entropy per
baryon is a dimensionless function of the dimensionless quantity
$T/\mu_{\rm B}$ and hence different QGP scenarios with the same
$T/\mu_{\rm B}$ all have the same value of the entropy per baryon.

As the result of this study we conclude that the charge flow ratio
with the charged particle multiplicity is an excellent experimental
measure of the entropy per baryon in the source, and that the current
results are suggestive of a much more entropy-rich emitter at central
rapidity than can be expected in hadronic gas models.

\pagebreak[5]

$\ $

\vspace{-2.2cm}
\hspace{2.6cm} fireball with $T \simeq$~200--220~MeV \cite{sollfrank} or,
including a transverse flow component, by $T \simeq 150$ MeV combined
with a flow velocity $\beta_{\rm f} \simeq 0.3 $ \cite{SH92}.

Following Eq.~(\ref{ratio}), the asymmetric $\overline{\Lambda} /
\Lambda$ ratio implies a non-zero combination of chemical potentials,
$\mu_{\rm B}+ (3/2)\,\mu_{\rm s} \simeq (0.8 \pm 0.4)\, T$. We note
that the experimental ratio decreases when averaged over a larger
rapidity interval, implying even larger chemical potentials. If
$T\simeq 210$ MeV, both the HG and QGP model imply $\mu_{\rm s} \simeq
0$ for strangeness neutral systems, and we thus obtain $\mu_{\rm B}
\simeq 170 \pm 85$ MeV.
For the ``flow" scenario with $T=150$ MeV,
strangeness conservation in a HG allows the above relation to be
satisfied with $\mu_{\rm s} \simeq 17$ MeV at $\mu_{\rm B}\simeq 95$
MeV, again with rather large experimental errors.

For the ``thermal" case we can use Fig.~\ref{F6} to translate
$\mu_{\rm B}$ into $D_{\rm Q}$; with $\mu_{\rm B} = 170 \pm 85$ MeV we
obtain $D_{\rm Q}\simeq 0.13\pm0.06$. Although the central value of
$D_{\rm Q}$ extracted in this way is twice the measured value for
this ratio, due to the large experimental uncertainty of the
$\overline{\Lambda} / \Lambda$ ratio the two numbers still agree
within one standard deviation. We note that a HG with $D_{\rm Q}=0.13$
possesses a specific entropy $\cal S/B$ = 35, while $D_{\rm Q}$ =
0.065 requires $\cal S/B$ = 65--75. For the ``flow" scenario, on the
other hand, we compute with the conventional HG-EoS (with $\mu_{\rm B}
= 95$ MeV, $\mu_{\rm s} = 17$ MeV) a value $D_{\rm Q}\simeq 0.05$,
consistent with the measured value 0.065.

As this discussion demonstrates, the current NA35 strange particle
data are neither precise enough nor do they provide sufficiently
complete information to allow for a similar analysis as done for the
WA85 data above. In particular the lack of data on $\Xi^- $ and
$\overline{\Xi^- }$ production in S--S collisions prevents us from
determining the values of $\gamma_{\rm s}$ and $\mu_{\rm s}$. In the
absence of such information, the validity of a HG description for the
NA35 S--S data can at the present moment not be excluded.

\section{Implications for the fireball evolution dynamics}
\label{sect5}

So far we have extracted, with the help of thermodynamical concepts,
from various sets of nuclear collision data the characteristic
statistical properties of the source of the emitted particles. In the
following we will comment on what our findings imply for the possible
formation of a quark-gluon plasma, for its evolution and ultimately its
dissociation into the final state hadrons.

We have seen that a HG interpretation of the WA85 data on strange
baryon production is not consistent with the large multiplicity
densities observed by EMU05 which indicate a much larger specific
entropy than implied by the HG picture. In this context we have
discussed both a ``purely thermal" interpretation of the data (which
identifies the slope of the transverse mass spectra directly as the
fireball temperature) and a ``flow scenario", where part of the
observed slope is due to a blueshift factor caused by transverse
collective expansion of the fireball. Independent of which version of
the HG picture we used, we found from the WA85 data a surprisingly
large degree of strangeness saturation $\gamma_{\rm s}$, ranging from
75\% in the purely ``purely thermal" interpretation to 90\% in the
``flow scenario". These are hard to understand in a HG scenario,
because the time scale for strangeness producing processes is usually
believed to be much too slow to achieve that kind of saturation during
the lifetime of the collision, unless a QGP is formed.

Within the HG picture, the dynamical aspects connected with the flow
seem to be required for consistency with the freeze-out kinetics
\cite{SH92}. For this reason the ``flow scenario" has been highly
favored before. However, we have shown here that within the HG
picture the value $\mu_{\rm s}=0$ constrains the liberty to interpret
the slope of the transverse mass spectra in terms of a temperature
which is blueshifted by flow. The observed values $\mu_{\rm s}=0$ and
$\mu_{\rm B}/T=1.63$ are inconsistent with strangeness neutrality of
the observed system of hadrons at the rather low value of $T\simeq
150$ MeV resulting from the ``flow'' picture. To recover consistency
with the strangeness neutrality of the fireball under these
conditions, the HG equation of state would have to be very different
from the conventional Hagedorn resonance gas which we employed here.
While such a major modification of the physics of the hadron phase
cannot at present be completely excluded, it would invalidate a large
body of successful phenomenological work based on the conventional HG
picture. Thus, if the experimental result $\mu_{\rm s}=0$ should be
confirmed via other particle ratios, in our opinion it practically
eliminates within the framework of a conventional HG-EoS the flow
interpretation of the spectra as a viable alternative to the purely
``thermal" interpretation.

We therefore conclude that within a HG picture the interpretation of
the $m_\bot$-slope of $\simeq210$ MeV as the true temperature of the
fireball appears to be the only possibility consistent with the
constraint of (approximate) strangeness neutrality. This excludes the
possibility of transverse flow, however, and thus causes consistency
problems for the freeze-out process. In addition, it does not
lead to the observed value of $D_{\rm Q}$. Therefore a HG
interpretation of the data appears to be excluded.

If one tries instead to interpret the observations in terms of QGP
formation, one faces the problem to understand how the hadronization
process can lead to a final state with the observed values for the
thermodynamic parameters. We have seen that the conditions of
strangeness neutrality and of a fixed up-down asymmetry in general
lead, at the same temperature, to different values for the chemical
potentials $\mu_{\rm u}$, $\mu_{\rm d}$, $\mu_{\rm s}$ in the two
phases. While a vanishing value of $\mu_{\rm s}$ is a natural
consequence of strangeness neutrality in the QGP, it is not at all
natural in the hadron gas, where $\mu_{\rm s}=0$ solves the
strangeness neutrality condition only for very specific pairs of $T$
and $\mu_{\rm B}$, and thus in general it is not a value expected to
arise in an interpretation of the experiments involving HG fireballs.
On the other hand, while the values of the chemical potentials
assumed in a hadron gas can be directly determined from the measured
abundance ratios between the various hadronic species emitted by the
gas, this is not automatically true for the QGP chemical potentials.
How $\mu_{\rm u}$, $\mu_{\rm d}$, and $\mu_{\rm s}$ inside the QGP
are related to the abundance ratios of the finally observed hadrons
depends on the dynamical and kinetic details of the hadronization
process.

To illustrate this problem, let us consider the limiting case of a
very long-lived and slowly hadronizing quark-gluon plasma, where
hadronization proceeds through a mixed phase providing intimate
contact between the two sub-phases. In this case chemical equilibrium
not only within the hadronic sub-phase (with respect to processes
which transform the hadrons among each other) and within the quark
sub-phase (with respect to processes transforming quarks and gluons
into each other), but also between the two phases (with respect to
processes which create hadrons out of quarks and gluons or dissociate
them again into quarks and gluons) can be maintained. As a
consequence, the chemical potentials change smoothly from their
values originally assumed in the QGP phase into the chemical
equilibrium values of the hadron gas, and only the latter (and not
the former) would be observable by studying particle ratios. In such
a scenario, the value of the strange quark chemical potential
$\mu_{\rm s}$ extracted from a suitable combination of strange hadron
abundances \cite{Raf91} would in general not be zero (although it
started out as zero in the original QGP phase), but would be given by
the solution to the strangeness balance constraint (\ref{Eq6}) in an
equilibrated hadron gas.

However, as demonstrated in this paper, the strange particle yields
from the WA85 collaboration \cite{WA85} indeed do point to a value of
$\mu_{\rm s}=0$, and in view of the naturalness of this value in the
context of a QGP we are inclined to believe that this result may not
be entirely accidental. Therefore we ask under which conditions
$\mu_{\rm s}=0$ could have survived the QGP hadronization process
after all.

One possibility to transfer the QGP values for the chemical potentials
and temperature directly to the final state hadrons is to assume rapid
break-up of the QGP combined with statistical quark recombination
\cite{Raf91,KMR86,BZ83,RD87}. A necessary condition for the
preservation of the QGP thermodynamic parameters is that sequentially
produced hadrons escape sufficiently rapidly such that there is no
formation of a intermediate HG phase in which the chemical potentials
and/or momentum distributions could re-equilibrate. One can view this
process as the peeling of the QGP surface into the vacuum --- since
there is no barrier, the surface regions of the QGP fireball are
peeled off in form of rapidly escaping hadrons, carrying in their
abundances the information about the thermal and chemical properties
of the QGP.

One problem with such a picture is that the system of
radiated hadrons, described by thermal distributions with chemical
potentials and temperature as given originally in the QGP phase,
by definition is characterized by the same particle abundance ratios
and by the same value of the specific entropy as an equilibrium HG at
the same temperature. The radiation obtained from this quark
recombination picture is in general neither strangeness
neutral\footnote{We thank H.  Satz for bringing
this point to our attention.}, nor does it carry away the correct
amount of entropy per baryon (as fixed by the original QGP)
It is well known that the entropy balance can be saved by considering
the entropy contained in the gluons of the QGP, letting them materialize
as hadrons via gluon fragmentation into additional quark-anti-quark
pairs. In \cite{KMR86} the additional quarks from gluon
fragmentation where hadronized according to the same laws of
combinatorics as the thermal quarks from the QGP; as a result the QGP
chemical potentials were not preserved, but the quark chemical
potentials were modified during hadronization in the direction of
their HG equilibrium values.

The minimal dynamical picture could consist of a
primordial explosive production of the few high $m_\bot$ strange anti-baryons,
followed by equilibrium hadronization of the remaining fireball
at nearly constant entropy.
To test such a picture requires the simultaneous measurement of the
abundances and momentum spectra of (strange and non-strange) baryons
and anti-baryons and of (fully identified) pions and kaons. Only with
such a complete set of data on produced hadrons can a full understanding
of the evolution of the primordial dense phase (QGP?) be achieved.

Of course, our contention that the presently available data are
inconsistent with a HG picture for the collision and rather point to a
very entropy-rich source of particle emission, possibly a quark-gluon
plasma, must appear provocative at the present stage. Our findings
show that it is necessary to widen the investigation of strange
anti-baryon ratios to higher {\em and\/} lower nuclear beam energies
and to study simultaneously the particle multiplicity and charge flow
at high and low $m_\bot$, in order to confirm the now very suggestive
conclusion of QGP formation in the collisions of S ions at 200 GeV A
with heavy nuclear targets at near zero impact parameter. In view of
the ambiguity between the HG and QGP picture, as far as the
consistency of a vanishing strange quark chemical potential with the
condition of strangeness neutrality is concerned, the presently
available beam energies at CERN turn out to be somewhat unfortunate,
due to the observed particular value of around 210 MeV for the
$m_\perp$-slope; this ambiguity could be easily resolved if $\mu_{\rm
s}=0$ were still observed in cases where the $m_\perp$-slope (apparent
temperature) is lower or higher. It is natural to expect such a
variation in the slope parameters (combined by an inverse variation of
the baryon chemical potential) once collisions at lower or higher beam
energies are studied (for example with a 50 GeV A Pb-beam at the CERN
SPS or with 100+100 GeV A Au-beams at RHIC). We are also looking forward
with anticipation to the forthcoming Pb--Pb collisions at about 170
GeV A projectile energy, which will allow to check and extend the
results of our analysis by going to the largest possible thermodynamic
systems using nuclear collisions.

\vskip 0.4cm

\noindent {\bf Acknowledgements:} J.R. thanks his colleagues in
Paris and Regensburg for their warm and personal hospitality. The
work of U.H. and J.S. was supported by DFG, BMFT and GSI. The
collaboration between U.H. and J.R. was supported by NATO
Collaborative Research Grant 910991. J.R. acknowledges funding of his
research by DOE, grant DE-FG02-92ER40733.

\vfill\eject


\end{document}